\documentclass[12pt]{article}
\usepackage{amssymb,bm,amsmath}
\usepackage{epsfig}

\renewcommand{\thefootnote}{\fnsymbol{footnote}}
\newcommand{\vev}[1]{{\langle{#1}\rangle}}
\newcommand{\mat}[1]{\begin{pmatrix} #1\end{pmatrix}}

\begin{document}

\title{
\begin{flushright}
\begin{minipage}{0.2\linewidth}
\normalsize
%arXiv:YYMM.NNNN \\
WU-HEP-13-02 \\
KUNS-2454 \\*[50pt]
\end{minipage}
\end{flushright}
{\Large \bf 
Flavor landscape of 10D SYM theory \\
with magnetized extra dimensions
\\*[20pt]}}

\author{
Hiroyuki~Abe$^{1,}$\footnote{
E-mail address: abe@waseda.jp}, \ 
Tatsuo~Kobayashi$^{2,}$\footnote{
E-mail address: kobayash@gauge.scphys.kyoto-u.ac.jp}, \
Hiroshi~Ohki$^{3,}$\footnote{
E-mail address: ohki@kmi.nagoya-u.ac.jp}, \
Keigo~Sumita$^{1,}$\footnote{
E-mail address: k.sumita@moegi.waseda.jp}
\\ and \ 
Yoshiyuki~Tatsuta$^{1,}$\footnote{
E-mail address: y\_tatsuta@akane.waseda.jp}\\*[20pt]
$^1${\it \normalsize 
Department of Physics, Waseda University, 
Tokyo 169-8555, Japan} \\
$^2${\it \normalsize 
Department of Physics, Kyoto University, 
Kyoto 606-8502, Japan} \\
$^3${\it \normalsize 
Kobayashi-Maskawa Institute for the 
Origin of Particles and the Universe (KMI),} \\
{\it \normalsize 
Nagoya University, Nagoya 464-8602, Japan} \\*[50pt]}

\date{
\centerline{\small \bf Abstract}
\begin{minipage}{0.9\linewidth}
\medskip 
\medskip 
\small
We study the flavor landscape of particle physics models based on a 
ten-dimensional  super Yang-Mills  theory compactified 
on magnetized tori preserving four-dimensional ${\cal N}=1$ supersymmetry. 
Recently, we constructed a semi-realistic model which contains 
the minimal supersymmetric standard model (MSSM) 
using an Ansatz of magnetic fluxes and orbifolding projections. 
However, we can consider more various configurations of magnetic fluxes and 
orbifolding projections preserving four-dimensional ${\cal N}=1$ 
supersymmetry. 
%here may be other possibilities to obtain models 
%hich contains the MSSM. 
We research systematically such possibilities for leading to MSSM-like models
and study their 
phenomenological aspects.
\end{minipage}
}

\begin{titlepage}
\maketitle
\thispagestyle{empty}
\clearpage
\tableofcontents
\thispagestyle{empty}
\end{titlepage}

\renewcommand{\thefootnote}{\arabic{footnote}}
\setcounter{footnote}{0}

\section{Introduction}
The standard model (SM) of elementary particles is a quite successful 
theory, consistent with all the experimental data obtained so far 
with a great accuracy. However, it does not describe gravitational 
interactions of 
elementary particles that could play an important role 
at the very beginning of our universe. 
Superstring theories in ten-dimensional (10D) spacetime are almost 
the only known candidates for ultimate unification theory of 
elementary particles including gravitational interactions. 
Supersymmetric Yang-Mills (SYM) theories 
in various spacetime dimensions appear as low energy effective 
theories of superstring compactifications with or without D-branes. 
That is quite attractive from the phenomenological viewpoint as well as 
the theoretical viewpoint.
Indeed, based on SYM effective theories, 
 various phenomenological aspects of superstring models 
have been studied so far. (See for a review \cite{Ibanez:2012zz}.)
It is an interesting possibility that the SM is embedded in 
one of such SYM theories, that is, the SM is realized as a low 
energy effective theory of the superstring theories. 
In such a string model building, it is a key issue 
how to break higher-dimensional 
supersymmetry and to obtain a four-dimensional (4D) chiral spectrum. 
String compactifications on the Calabi-Yau (CY) space provide a 
general procedure for such a purpose. However, the metric of 
a generic CY space is difficult to be determined analytically.
That makes phenomenological studies qualitative, 
but not quantitative.

It is quite interesting that even simple toroidal compactifications 
but with magnetic fluxes in extra dimensions induce 4D chiral 
spectra starting with higher-dimensional SYM theories 
as well as superstring theories~\cite{Bachas:1995ik,Cremades:2004wa}. 
The higher-dimensional supersymmetry such as ${\cal N}=4$ in terms 
of supercharges in 4D spacetime is broken by 
magnetic fluxes down to 4D ${\cal N}=0$, $1$ or $2$ depending on the 
configuration of fluxes. The number of the chiral zero-modes is 
determined by the magnitude of magnetic fluxes. A phenomenologically 
attractive feature is that these chiral zero modes localize at 
different points in magnetized extra dimensions. 
The overlap integrals of localized wavefunctions yield hierarchical 
couplings in the 4D effective theory of these zero modes. 
That could explain, e.g., observed hierarchical masses and mixing angles 
of the quarks and the leptons~\cite{Abe:2008sx}. 
Furthermore, higher-order couplings can also be computed as 
the overlap integrals of wavefunctions~\cite{Abe:2009dr}.
A theoretically attractive point here is that many peculiar 
properties of the SM, such as the 4D chirality, the number 
of generations, the flavor symmetries~\cite{Abe:2009vi,Abe:2009uz,BerasaluceGonzalez:2012vb,Marchesano:2013ega} 
and potentially hierarchical Yukawa couplings 
all could be determined by the magnetic fluxes.

%Moreover if the 4D ${\cal N}=1$ supersymmetry remains, 
%a supersymmetric standard model could be realized below the 
%compactification scale that has many attractive features 
%beyond the SM, such as the lightest supersymmetric particle 
%as a candidate of dark matter and so on. 

In our previous works~\cite{Abe:2012ya,Abe:2012fj}, we have presented 
4D ${\cal N}=1$ superfield description of 10D SYM theories 
compactified on magnetized tori which preserve the 
${\cal N}=1$ supersymmetry, and derived 4D effective action 
for massless zero-modes written in the ${\cal N}=1$ superspace. 
It makes easy to construct models and analysis that. 
Thanks to that, we have constructed a three-generation model of quark 
and lepton chiral superfields based on a toroidal compactification 
of the 10D SYM theory with certain magnetic fluxes in extra
dimensions preserving a 4D ${\cal N}=1$ supersymmetry. 
The low-energy effective theory contains the
minimal supersymmetric standard model (MSSM) 
particle contents, where the three generation structure of chiral matter 
fields is originated by the magnitudes 
of fluxes, and it gives a semi-realistic pattern of
hierarchical masses and mixing angles of the quarks and the
leptons~\cite{Abe:2012fj}. 
Furthermore, we estimated the size of supersymmetric flavor violations.

%We further identified moduli dependence of the effective 
%action by promoting the Yang-Mills (YM) gauge coupling constant 
%$g$ and geometric parameters $R_i$ and $\tau_i$ to a dilaton, 
%K\"ahler and complex-structure moduli superfields, which 
%allows an explicit estimation of soft supersymmetry breaking 
%parameters in the supersymmetric SM caused by moduli-mediated 
%supersymmetry breaking. 
%The resulting effective supergravity action would be useful 
%for building phenomenological models and for analyzing them 
%systematically. 

One typical model has been studied in our previous works, however, 
there may be other possibilities for particle physics models 
based on a 10D SYM theory compactified on three factorizable tori 
$T^2 \times T^2 \times T^2$ where magnetic fluxes are present 
in the YM sector. In this paper, we search systematically such possibilities 
to obtain models which contains the MSSM 
and study the flavor landscape of 10D magnetized SYM theories, 
requiring full rank Yukawa matrices among three generations of 
quarks and leptons. 
We include possibilities of non-factorizable 
magnetic
fluxes~\cite{Cremades:2004wa,Antoniadis:2009bg,DeAngelis:2012jc}.
(See also \cite{Sakamoto:2003rh}.)
We also consider the $T^6/Z_2$ orbifold with magnetic fluxes 
as the geometrical background \cite{Abe:2008fi}.\footnote{Recently 
the orbifold compactification with magnetic fluxes was studied within 
the framework of heterotic string theory 
\cite{Nibbelink:2012de}.}$^{,}$\footnote{Here, we consider the twist
orbifolding, although shift orbifolding is also possible 
\cite{Fujimoto:2013xha}.}

The sections are organized as follows. 
In Sec.~\ref{sec:10d}, a superfield description of the 10D SYM 
theory is briefly reviewed based on Ref.~\cite{Abe:2012ya}, 
which allows the systematic introduction of magnetic fluxes 
in extra dimensions preserving the ${\cal N}=1$ supersymmetry. 
Then, we introduce (factorizable) magnetic fluxes and search 
systematically their proper configurations to construct models 
that contains the spectrum of the MSSM with or without orbifold 
projection in Sec.~\ref{sec:mag}. 
In Sec.~\ref{sec:obl}, we consider magnetic fluxes which cross over 
two tori. It gives possibilities of more various model building to us. 
Including such nonfactorizable fluxes, we carry out a
 systematic search with or without 
orbifold projection. Sec.~\ref{sec:con} is devoted to conclusions and 
discussions. 
In Appendix A, we show one $Z_2$ orbifold model, where 
most of exotic zero-modes are projected out.
In Appendix B, we show explicitly Yukawa matrices in one
type1 model shown in Sec.~\ref{sec:s2hd}

\section{The 10D SYM theory in ${\cal N}=1$ superspace}
\label{sec:10d}

We consider a compactification of 10D SYM 
theory on 4D flat  Minkowski spacetime times a product of 
factorizable three tori $T^2 \times T^2 \times T^2$ 
and use a superfield description suitable for such a 
compactification with magnetic fluxes 
preserving 4D ${\cal N}=1$ supersymmetry. 
Basically, we assume that the compactification scale is around 
the GUT or Planck scale, although application to other
compactifications  is straightforward.
We mostly follow the notations and the conventions 
adopted in our previous works \cite{Abe:2012ya,Abe:2012fj}.
Here, we briefly review  
only essential points for this work. 

%The 10D SYM theory is described by the following action 
%\begin{eqnarray}
%S &=& \int d^{10}X\,\sqrt{-G}\,
%\frac{1}{g^2}\,{\rm Tr}\left[ 
%-\frac{1}{4}F^{MN}F_{MN} 
%+\frac{i}{2} \bar\lambda \Gamma^M D_M \lambda 
%\right], 
%\label{eq:10dsym}
%\end{eqnarray}
%where $M,N=0,1,\ldots,9$. 

We decompose the 10D vector (gauge) field $A_M$ into 
the 4D vector field $A_\mu$ and three complex fields $A_i$ $(i=1,2,3)$. 
Also the 10D Majorana-Weyl spinor field $\lambda$ is decomposed 
into four 4D Weyl spinor fields $\lambda_0$ and $\lambda_i$ $(i=1,2,3)$. 
For 4D positive chirality, these spinor fields 
$\lambda_0$, $\lambda_1$, $\lambda_2$, and $\lambda_3$ 
have the 6D chiralities, 
$(+,+,+)$, $(+,-,-)$, $(-,+,-)$, and $(-,-,+)$ on 
$T^2 \times T^2 \times T^2$. 
The 10D SYM theory possesses ${\cal N}=4$ supersymmetry 
counted in terms of 4D supercharges. The YM vector and spinor 
fields, $A_M$ and $\lambda$, are decomposed into (on-shell) 
4D ${\cal N}=1$ single vector and triple chiral multiplets, 
${\bm V}=\left\{ A_\mu, \lambda_0 \right\}$ and 
${\bm \phi}_i \ = \ \left\{ A_i, \lambda_i \right\}$ $(i=1,2,3)$.

The above ${\cal N}=1$ vector and chiral multiplets, 
${\bm V}$ and ${\bm \phi}_i$, are expressed by vector and 
chiral superfields, $V$ and $\phi_i$, respectively. 
%\begin{eqnarray}
%V &\equiv& -\theta\sigma^\mu\bar\theta A_\mu
%+i\bar\theta\bar\theta\theta\lambda_0 
%-i\theta\theta\bar\theta\bar\lambda_0 
%+\frac12\theta\theta\bar\theta\bar\theta D, 
%\nonumber \\
%\phi_i &\equiv& \frac1{\sqrt2} A_i
%+\sqrt2\theta\lambda_i+\theta\theta F_i, 
%\nonumber
%\end{eqnarray}
%where $\theta$ and $\bar\theta$ are 
%Grassmann coordinates of 4D ${\cal N}=1$ superspace. 
Using them, the 10D SYM action can be written in the 
${\cal N}=1$ superspace as~\cite{Marcus:1983wb} 
\begin{eqnarray}
S &=& \int d^{10}X \sqrt{-G} \left[ \int d^4\theta\,{\cal K}
+\left\{ \int d^2 \theta\,\left( 
\frac{1}{4g^2} {\cal W}^\alpha {\cal W}_\alpha 
+ {\cal W} \right) +{\rm h.c.} \right\} 
\right]. 
\nonumber
\end{eqnarray}
The functions of the superfields, 
${\cal K}$, ${\cal W}$ and ${\cal W}_\alpha$, are given by 
\begin{eqnarray}
{\cal K} &=& \frac{2}{g^2}  h^{\bar{i}j} 
{\rm Tr} \left[\left(\sqrt2\bar\partial_{\bar{i}}
+\bar\phi_{\bar{i}}\right)e^{-V}
\left(-\sqrt2\partial_j + \phi_j \right)e^V
+\bar\partial_{\bar{i}} e^{-V} \partial_j e^V \right] 
+{\cal K}_{\rm WZW}, 
\nonumber \\
{\cal W} &=& \frac{1}{g^2} 
\epsilon^{{\rm i}{\rm j}{\rm k}} 
e_{{\rm i}}^{\ i} e_{{\rm j}}^{\ j} e_{{\rm k}}^{\ k} 
{\rm Tr} \left[ \sqrt{2}\, \phi_i
\left(\partial_j\phi_k-\frac{1}{3\sqrt2}
\left[\phi_j,\phi_k\right]\right)\right], 
\nonumber \\
{\cal W}_\alpha &=& 
-\frac{1}{4} \bar{D} \bar{D} e^{-V} D_\alpha e^V, 
\nonumber
\end{eqnarray}
where $\partial_i$ denotes a derivative of complex coordinate $z_i$, 
and $h^{\bar ij}$ and 
$e_{{\rm i}}^{\ i}$ are metric and vielbein of the $i$-th torus. 
The term ${\cal K}_{\rm WZW}$ denotes a 
Wess-Zumino-Witten term which vanishes 
in the Wess-Zumino (WZ) gauge. 
Note that only the combination $\phi_1 \phi_2 \phi_3$ has 
non-vanishing Yukawa couplings in ${\cal W}$.

The equations of motion for 
auxiliary fields $D$ and $F_i$ lead to 
\begin{eqnarray}
D &=& - h^{\bar{i}j} \left( 
\bar\partial_{\bar{i}} A_j + \partial_j \bar{A}_{\bar{i}} 
+\frac{1}{2} \left[ \bar{A}_{\bar{i}}, A_j \right] \right), 
\label{eq:osd} \\
\bar{F}_{\bar{i}} &=& -h_{j\bar{i}}\, 
\epsilon^{{\rm j}{\rm k}{\rm l}} 
e_{{\rm j}}^{\ j} e_{{\rm k}}^{\ k} e_{{\rm l}}^{\ l} 
\left( \partial_k A_l 
-\frac{1}{4} \left[ A_k,\, A_l \right] \right). 
\label{eq:osfi}
\end{eqnarray}
The condition 
$\langle D \rangle = \langle F_i \rangle =0$ 
determines supersymmetric vacua. 
A trivial supersymmetric vacuum is given by 
$\langle A_i \rangle = 0$ where the full ${\cal N}=4$ 
supersymmetry as well as the YM gauge symmetry is preserved. 
In the following, we search nontrivial supersymmetric 
vacua where magnetic fluxes exist in the YM sector.

\section{Magnetic fluxes and zero-modes}
\label{sec:mag}

We consider the 10D $U(N)$ SYM theory\footnote{A similar study is 
possible by starting with other gauge groups \cite{Choi:2009pv}.}
 on a supersymmetric background with factorizable magnetic fluxes. 
In this section, we study the case that the YM fields take the following 
4D Lorentz invariant and at least ${\cal N}=1$ 
supersymmetric vacuum expectation values (VEVs), 
\begin{eqnarray}
\vev{A_i} &=& \frac{\pi}{{\rm Im}\, \tau_i} 
\left( M^{(i)}\,\bar{z}_{\bar{i}}+\bar\zeta_i \right), \qquad 
\vev{A_\mu} \ = \ 
\vev{\lambda_0} \ = \ 
\vev{\lambda_i} \ = \ 
\vev{F_i} \ = \ 
\vev{D} \ = \ 0. 
\label{eq:ymvevs}
\end{eqnarray}
This VEV of ${A_i}$ leads to factorizable magnetic fluxes, 
and this form is unique up to gauge transformation for the fixed 
magnitude of $M^{(i)}$.
Here $M^{(i)}$ denotes the 
$N \times N$ diagonal matrices of Abelian magnetic fluxes 
as 
\begin{eqnarray}
M^{(i)} &=&\left( 
\begin{array}{cccc}
M^{(i)}_1 {\bm 1}_{N_1} & \ & \ & \ \\
\ & M^{(i)}_2 {\bm 1}_{N_2} & \ & \ \\
\ & \ & \ddots & \ \\
\ & \ & \ & M^{(i)}_n {\bm 1}_{N_n} \ \\
\end{array} \right), \label{eq:abelianflux} \\
\nonumber
\end{eqnarray}
where ${\bm 1}_{N_a}$ is a $N_a \times N_a$ unit matrix.
Similarly, the Wilson lines $\zeta_i$ are denoted as 
\begin{eqnarray*}
\zeta^{(i)} &=&\left( 
\begin{array}{cccc}
\zeta^{(i)}_1 {\bm 1}_{N_1} & \ & \ & \ \\
\ & \zeta^{(i)}_2 {\bm 1}_{N_2} & \ & \ \\
\ & \ & \ddots & \ \\
\ & \ & \ & \zeta^{(i)}_n {\bm 1}_{N_n} \ \\
\end{array} \right). \label{eq:WL} \\
\end{eqnarray*}
Here, $\tau_i$ is complex structure of
the $i$-th torus.
We restrict $\tau_i$ to be pure imaginary, 
since its real part only affects 
physical CP phases. 
The magnetic fluxes satisfying 
the Dirac's quantization condition, 
$M_1^{(i)},M_2^{(i)},\ldots,M_n^{(i)} \in \bm Z$, 
are further constrained by the supersymmetry conditions 
$\vev{D}=0$ and $\vev{F_i}=0$ in Eq.~(\ref{eq:ymvevs}), 
which are written as 
\begin{eqnarray}
h^{\bar{i}j} \left( \bar\partial_{\bar{i}} \vev{A_j} 
+ \partial_j \vev{\bar{A}_{\bar{i}}} \right) &=& 0, 
\label{eq:abd} \\
\epsilon^{{\rm j}{\rm k}{\rm l}} 
e_{{\rm k}}^{\ k} e_{{\rm l}}^{\ l} 
\partial_k \vev{A_l} &=& 0,
\label{eq:abfi}
\end{eqnarray}
with $D$ and $F_i$ given by Eq.~(\ref{eq:osd}) 
and (\ref{eq:osfi}), respectively. Magnetic fluxes with the VEVs 
Eq.~(\ref{eq:ymvevs}) trivially satisfy the F term condition 
Eq.~(\ref{eq:abfi}). 

%One of the consequences of nonvanishing magnetic fluxes is 
%the YM gauge symmetry breaking. 
The magnetic flux (\ref{eq:abelianflux}) 
breaks the $U(N)$ gauge symmetry into $\prod_a U(N_a)$ for $a=1,2,\ldots,n$ 
if all the magnetic fluxes $M_1^{(i)},M_2^{(i)},\ldots,M_n^{(i)}$ 
take values different from each other. The same holds for Wilson-lines, 
$\zeta_1^{(i)},\zeta_2^{(i)},\ldots,\zeta_N^{(i)}$. 
In the following, the indices $a,b=1,2,\ldots,n$ 
label the unbroken YM subgroups on the flux and Wilson-line 
background~(\ref{eq:ymvevs}), and traces in expressions 
are performed within such subgroups.

On the ${\cal N}=1$ supersymmetric toroidal background~(\ref{eq:ymvevs}) 
with the magnetic fluxes~(\ref{eq:abelianflux}) as well as the Wilson-lines, 
the zero-modes $(V^{{\bm n}={\bm 0}})_{ab}$ 
of the off-diagonal elements $(V)_{ab}$ ($a \ne b$) 
in the 10D vector superfield $V$ obtain mass terms, 
while the diagonal elements 
$(V^{{\bm n}={\bm 0}})_{aa}$ do not. 
Then, we express the zero-modes 
$(V^{{\bm n}={\bm 0}})_{aa}$, 
which contain 4D gauge fields for the unbroken gauge symmetry 
$\prod_a U(N_a)$, as 
\begin{eqnarray}
(V^{{\bm n}={\bm 0}})_{aa} &\equiv& V^a. 
\nonumber
\end{eqnarray}
The magnetic fluxes and the Wilson-lines have no effect on 
these diagonal elements, and extra dimensional parts of 
their wavefunctions are constant. 
On the other hand, for $^\exists j \ne i$ with 
$M^{(j)}_{ab} \equiv M^{(j)}_a - M^{(j)}_b > 0$, 
$M^{(i)}_{ab} \equiv M^{(i)}_a - M^{(i)}_b < 0$ 
and $a \ne b$, 
the zero-mode $(\phi_j^{{\bm n}={\bm 0}})_{ab}$ 
of the off-diagonal element $(\phi_j)_{ab}$ 
in the 10D chiral superfield $\phi_j$ 
degenerates with the number of degeneracy 
$N_{ab}=\prod_k \left| M^{(k)}_{ab} \right|$, 
while $(\phi_j^{{\bm n}={\bm 0}})_{ba}$ 
has no zero-mode solution, 
yielding a 4D supersymmetric chiral generation 
in the $ab$-sector~\cite{Abe:2012ya}. 
%The opposite is true for 
%$M^{(j)}_{ab} > 0$ and $M^{(i)}_{ab} < 0$ 
%yielding a 4D chiral generation in the $ba$-sector. 
Therefore, we denote the zero-mode 
$(\phi_j^{{\bm n}={\bm 0}})_{ab}$ 
with the degeneracy $N_{ab}$ as 
\begin{eqnarray}
(\phi_j^{{\bm n}={\bm 0}})_{ab} &\equiv& 
g\,\phi_j^{{\cal I}_{ab},M^{(i)}_{ab}}, 
\nonumber
\end{eqnarray}
where 
${\cal I}_{ab}$ labels the degeneracy, i.e. generations. 
We normalize $\phi_j^{{\cal I}_{ab}}$ by 
the 10D YM coupling constant $g$. 
An extra dimensional part of their wavefunctions is decomposed 
into three parts and each of them corresponds to the $i$-th torus.
These wavefunctions can be written by the 
Jacobi theta-function. For example, for the $i$-th torus, 
the zero-mode wavefunctions are written by 
\begin{gather}
\phi_j^{{\cal I}^{(i)}_{ab},M^{(i)}_{ab}}(\tau_i, z_i)=\mathcal N \cdot e^{i \pi M^{(i)}_{ab}z_i 
\mathrm{Im}\, z_i / \mathrm{Im}\, \tau_i} \cdot \vartheta \begin{bmatrix}
{\cal I}^{(i)}_{ab}/{M^{(i)}_{ab}}\\[5pt] 0\end{bmatrix} (M^{(i)}_{ab}z_i, M^{(i)}_{ab}\tau_i),\label{eq:zerowave}
\end{gather}
where ${\cal I}^{(i)}_{ab}$ labels the degeneracy generated on the $i$-th 
torus and $\vartheta$ is the Jacobi theta-function, 
\begin{gather}
\vartheta \begin{bmatrix}a\\[5pt] b\end{bmatrix} (\nu, \tau) = \sum_{l \in \mathbb{Z}} e^{\pi i (a+l)^2 \tau} e^{2 \pi i (a+l) (\nu+b)}.\label{eq:jacobi}
\end{gather}
The normalization factor $\mathcal N$ is defined as follows
\begin{gather}
\int \limits_{T^2} dzd\bar{z} \, \phi_j^{I,M}(\phi_j^{K,M})^*
=\delta_{IK} .
\nonumber
\end{gather}
For more details, see Ref.~\cite{Abe:2012ya} 
and references therein.

\subsection{Factorizable flux Ansatz}
%From now on, we call the flux $M^{(i)}$ in $\vev{A_i}$, 
%like (\ref{eq:abelianflux}), factorizable fluxes. 
We aim to realize a zero-mode spectrum in 10D SYM theory 
compactified on magnetized tori, that contains the MSSM with 
the gauge symmetry $SU(3)_C \times SU(2)_L \times U(1)_Y$ and 
three generations of the quark and the lepton chiral multiplets, 
by identifying these three generations with degenerate zero-modes of 
the chiral superfields $\phi_j^{{\cal I}_{ab}}$. 

To obtain the SM gauge group, especially $U(1)_Y$, we have to start 
at least the 10D U(8) SYM theory which contains the Pati-Salam 
gauge group, $U(4)_C \times U(2)_L \times U(2)_R$, up to 
$U(1)$ factors.
In the Pati-Salam gauge group, 
$(4,2,1)$  matter fields correspond to 
left-handed quarks and leptons, while 
$(\bar 4,1,2)$  matter fields correspond to 
right-handed quarks and leptons.
In addition, Higgs fields are assigned to $(1,2,2)$ representations.
%We can consider the SYM theory with a lager gauge group, 
%and if one aim to some extended MSSM it is very attractive. 
%However,  lager gauge groups generally lead various 
%massless exotic modes and they couple with the MSSM sector. 
%At any rate, we consider only U(8) gauge group in this paper.
Although we can start a larger gauge group, we concentrate ourselves 
on the U(8) gauge group as our starting point.

In our models, magnetic fluxes yield potentially hierarchical 
Yukawa couplings from 10D gauge couplings, the tri-linear terms 
of $\mathcal W$.
We require full-rank Yukawa matrices among three generations.
%Although we will not estimate exact values of Yukawa 
%matrices in this paper, we note the following. 
If the flavor structure of left-handed and right-handed matter 
fields are originated from different tori, one obtains rank-one 
Yukawa matrices at the tree-level. It leads two massless generations 
(at the tree-level).
That cannot be realistic unless there are corrections such as 
non-perturbative effects.
We will derive three generation structures 
of all the chiral matter fields on the first torus to obtain full-rank Yukawa 
matrices.
%(If we obtain the flavor structures on the second or the third 
%torus, it gives the same landscape).

Now, we study proper configurations of Abelian factorizable magnetic fluxes 
Eq.~(\ref{eq:ymvevs}) and (\ref{eq:abelianflux}). To obtain three generation 
structures on the first torus, we introduce fluxes 
with $M^{(1)}_{ab}=3$ for the quark and lepton chiral multiplets. 
The simplest one is 
\begin{eqnarray}
M^{(1)} &=& \left( 
\begin{array}{cc}
0 \times {\bm 1}_4 &  \ \\
\ & 3 \times {\bm 1}_4 
\end{array} \right).\label{eq:twoblock}
\end{eqnarray}
Recall that $(4,2,1)$ and $(\bar 4,1,2)$ correspond to quarks and leptons.
Since only differences between magnetic fluxes have effects on physics, 
we fix the first element along the $U(1)$ direction of $U(4)_C$ to be vanishing.
We set  $M^{(2)}$ and $M^{(3)}$ such that they  
do not generate extra generations and satisfy 
SUSY preserving conditions Eq.~(\ref{eq:abd}) and (\ref{eq:abfi}). 
In addition, we consider the gauge groups to be further broken down to 
$U(3)_C \times U(1)_{C'} \times U(2)_L 
\times U(1)_{R'} \times U(1)_{R''}$ 
by the Wilson-lines,
\begin{eqnarray}
\zeta_1 &=& \left( 
\begin{array}{ccccc}
\zeta^{(1)}_C {\bm 1}_3 & \ & \ & \ & \ \\
\ & \zeta^{(1)}_{C'} & \ & \ & \ \\
\ & \ & \zeta^{(1)}_L {\bm 1}_2 & \ & \ \\
\ & \ & \ & \zeta^{(1)}_{R'} & \ \\
\ & \ & \ & \ & \zeta^{(1)}_{R''} 
\end{array} \right) ,
\label{eq:wlm}
\end{eqnarray}
where 
all the nonvanishing entries take values different from 
each other. 
However, in this two-block Ansatz (\ref{eq:twoblock}), 
the Higgs fields with the $(1,2,2)$ representation have 
no effects due to magnetic fluxes, 
and their zero-mode profiles are just flat.
Thus, the Yukawa matrices are proportional to 
the $(3\times 3)$ unit matrix and that is not realistic.\footnote{
By introducing Wilson lines, the zero-mode profiles of 
the Higgs fields become non-trivial and Yukawa couplings 
may be modified \cite{Hamada:2012wj}.
However, such Wilson lines make the Higgs fields massive, 
and the Wilson lines corresponding to $\cal {O}$(100) GeV mass scale 
would not lead sufficient modification to lead realistic results. 
}
To construct the plausible models which have three generation matter fields and full-rank Yukawa matrices, more complicated fluxes are needed.

The next-to-simplest one is as follows,
\begin{eqnarray}
M^{(1)} &=& \left( 
\begin{array}{ccc}
0 \times {\bm 1}_4 & \ & \ \\
\ & 3 \times {\bm 1}_2 & \ \\
\ & \ & -3 \times {\bm 1}_2 
\end{array} \right).\label{eq:pati}
\end{eqnarray}
We also set $M^{(2)}$ and $M^{(3)}$ such that they  
do not generate extra generations and satisfy 
SUSY preserving conditions Eq.~(\ref{eq:abd}) and (\ref{eq:abfi}). 
The gauge group is broken as $U(8) \to U(4)_C \times U(2)_L \times U(2)_R$, 
the Pati-Salam gauge group, which is broken 
by the Wilson-lines Eq.~(\ref{eq:wlm}) in a similar way. 
This type of magnetic fluxes, the three-block type, will be studied in detail 
in the following sections.

We can consider more complicated flux configurations such as 
the four-block or five-block magnetic fluxes, where 
$U(4)_C$ is broken as  $U(4)_C \rightarrow U(3)_C \times U(1)_C'$ 
and/or  $U(2)_R$ is broken as $U(2)_R \rightarrow U(1)_{R'} \times U(1)_{R''}$ 
by magnetic fluxes.
However, in those models one can not obtain full-rank Yukawa 
matrices satisfying the SUSY conditions.
Thus, we do not consider such possibilities, 
and we concentrate on the three-block Ansatz of magnetic fluxes 
(\ref{eq:pati}) with  the Wilson-lines Eq.~(\ref{eq:wlm}).

\subsection{Pati-Salam model}

We show the details of the Pati-Salam models in which the flavor structures 
are originated by the flux Eq.~(\ref{eq:pati}). The Pati-Salam gauge group 
generated by the flux is further broken down to 
$U(3)_C \times U(1)_{C'} \times U(2)_L 
\times U(1)_{R'} \times U(1)_{R''}$ 
by the Wilson-lines (\ref{eq:wlm}).
The gauge symmetries $SU(3)_C$ and $SU(2)_L$ 
of the MSSM are embedded into the above 
unbroken gauge groups as $SU(3)_C \subset U(3)_C$ and 
$SU(2)_L \subset U(2)_L$. 
The hypercharge $U(1)_Y$ is obtained as a proper linear 
combinations of U(1)'s.

We show all the possible patterns of magnetic flux configuration
satisfying 
SUSY preserving condition Eq.~(\ref{eq:abd}) and (\ref{eq:abfi}) as a 
result of systematic search in Table~\ref{tb:3models} 
by using the following notation:
\begin{eqnarray}
M^{(i)} &=& \left( 
\begin{array}{ccccc}
M^{(i)}_C {\bm 1}_4 & \ & \ \\
\ & M^{(i)}_L {\bm 1}_2 & \ \\
\ & \ & M^{(i)}_R  {\bm 1}_2 
\end{array} \right). \nonumber
\end{eqnarray}
There are three patterns and one of them, pattern 1, is studied 
in Ref.~\cite{Abe:2012fj} and these three lead 
phenomenological features quite similar  to each other. 
Thus, we show the details 
of the first one in the following. 

\begin{table}[htb]
\begin{center}
\begin{tabular}{|c|c|c|c|} \hline
&$(M^{(1)}_C,M^{(1)}_L,M^{(1)}_R)$ & $(M^{(2)}_C,M^{(2)}_L,M^{(2)}_R)$ & $(M^{(3)}_C,M^{(3)}_L,M^{(3)}_R)$ \\ \hline\hline
pattern 1&
$(0,+3,-3)$ & 
$(0,-1,0)$ & 
$(0,0,+1)$ 
\\ \hline\hline
pattern 2&
$(0,+3,-3)$ & 
$(0,-1,0)$ & 
$(0,+1,+1)$ 
\\ \hline\hline
pattern 3&
$(0,+3,-3)$ & 
$(0,-1,-1)$ & 
$(0,0,+1)$ 
\\ \hline
\end{tabular}
\end{center}
\caption{Three patterns to give the Pati-Salam model }
\label{tb:3models}
\end{table}

%We show more details of these models in Appendix \ref{sec:appa} and simply summ%arize here. 

%There are three types and one of them, type 1, is studied 
%in Ref.~\cite{Abe:2012fj} and these three lead quite similar 
%phenomenological features quite similar  to each other. 
%Thus, we show the details 
%of the first one in the following. 

The first one, which yields three generations of quarks and leptons 
and also the full-rank Yukawa matrices, satisfies the SUSY preserving 
conditions Eq.~(\ref{eq:abfi}) and (\ref{eq:abd}) with
\begin{gather}
{\cal A}^{(1)}/{\cal A}^{(2)}={\cal A}^{(1)}/{\cal A}^{(3)}=3,
\nonumber
\end{gather}
where $\mathcal A^{(i)}$ is an area of the $i$-th torus. 

In this configuration of magnetic fluxes, chiral superfields 
$Q$, $U$, $D$, $L$, $N$, $E$, $H_u$ and $H_d$ 
carrying 
the left-handed quark, 
the right-handed up-type quark, 
the right-handed down-type quark, 
the left-handed lepton, 
the right-handed neutrino, 
the right-handed electron, 
the up- and the down-type Higgs bosons, respectively, 
are found in $\phi_i^{{\cal I}_{ab}}$ as 
\begin{align}
\phi_1^{{\cal I}_{ab}} &= 
\left( 
\begin{array}{cc|c|cc}
\Omega_C^{(1)} & \Xi_{CC'}^{(1)} & 0 & 
\Xi_{CR'}^{(1)} & \Xi_{CR''}^{(1)} \\
\Xi_{C'C}^{(1)} & \Omega_{C'}^{(1)} & 0 & 
\Xi_{C'R'}^{(1)} & \Xi_{C'R''}^{(1)} \\ 
\hline 
\Xi_{LC}^{(1)} & \Xi_{LC'}^{(1)} & \Omega_L^{(1)} & 
H_u^K & H_d^K \\ 
\hline 
0 & 0 & 0 & \Omega_{R'}^{(1)} & \Xi_{R'R''}^{(1)} \\
0 & 0 & 0 & \Xi_{R''R'}^{(1)} & \Omega_{R''}^{(1)} 
\end{array}
\right), 
\nonumber \\
\phi_2^{{\cal I}_{ab}} &= 
\left( 
\begin{array}{cc|c|cc}
\Omega_C^{(2)} & \Xi_{CC'}^{(2)} & Q^I & 0 & 0 \\
\Xi_{C'C}^{(2)} & \Omega_{C'}^{(2)} & L^I & 0 & 0 \\
\hline 
0 & 0 & \Omega_L^{(2)} & 0 & 0 \\
\hline 
0 & 0 & 0 & \Omega_{R'}^{(2)} & \Xi_{R'R''}^{(2)} \\
0 & 0 & 0 & \Xi_{R''R'}^{(2)} & \Omega_{R''}^{(2)} 
\end{array}
\right), 
\nonumber \\
\phi_3^{{\cal I}_{ab}} &= 
\left( 
\begin{array}{cc|c|cc}
\Omega_C^{(3)} & \Xi_{CC'}^{(3)} & 0 & 0 & 0 \\
\Xi_{C'C}^{(3)} & \Omega_{C'}^{(3)} & 0 & 0 & 0 \\
\hline 
0 & 0 & \Omega_L^{(3)} & 0 & 0 \\
\hline 
U^J & N^J & 0 & \Omega_{R'}^{(3)} & \Xi_{R'R''}^{(3)} \\
D^J & E^J & 0 & \Xi_{R''R'}^{(3)} & \Omega_{R''}^{(3)} 
\end{array}
\right),\label{app.phis1} 
\end{align}
where the rows and the columns of matrices 
correspond to 
$a=C,C',L,R',R''$ and 
$b=C,C',L,R',R''$, 
respectively, and the indices $I,J=1,2,3$ and 
$K=1,\ldots,6$ label the zero-mode degeneracy, 
i.e., generations. 
Therefore, three generations of 
$Q$, $U$, $D$, $L$, $N$, $E$ and 
six pairs of $H_u$ and $H_d$ 
are generated by the pattern 1 magnetic fluxes in Table~\ref{tb:3models} 
that correspond to 
\begin{align}
M^{(1)}_C-M^{(1)}_L = -3, \qquad 
M^{(1)}_L-M^{(1)}_R &= +6, \qquad 
M^{(1)}_R-M^{(1)}_C = -3, \nonumber \\
M^{(2)}_C-M^{(2)}_L = +1, \qquad 
M^{(2)}_L-M^{(2)}_R &= -1, \qquad 
M^{(2)}_R-M^{(2)}_C = 0, \nonumber \\
M^{(3)}_C-M^{(3)}_L = 0, \qquad 
M^{(3)}_L-M^{(3)}_R &= -1, \qquad 
M^{(3)}_R-M^{(3)}_C = +1.\label{app.fluxes1} 
\end{align}
The zero entries of the matrices in Eq. (\ref{app.phis1}) 
represent components eliminated because of the 
chirality projection effects caused by magnetic fluxes. 
%Because some vanishing fluxes are inevitable in Eq.~(\ref{app.phis1}) 
In order to realize three generations of quarks and leptons 
with the full-rank Yukawa coupling matrices, some of 
$M^{(i)}_{ab}$ become vanishing in Eq.~(\ref{app.fluxes1}).
That causes certain massless exotic modes $\Xi_{ab}^{(r)}$ 
as well as massless diagonal components $\Omega_{a}^{(r)}$, 
i.e., the so-called open string moduli, all of which feel zero fluxes. 
These exotics are severely constrained by many experimental 
data at low energies. 
However, most of the massless exotic modes $\Xi_{ab}^{(r)}$ 
and diagonal components $\Omega_{a}^{(r)}$ 
can be projected out with the three generations of quarks and leptons unchanged 
if we impose further a certain orbifold projection 
on the second and third tori ~\cite{Abe:2012fj}.\footnote{
We show that in Appendix~\ref{sec:appa}.}

The three patterns shown in Table.~\ref{tb:3models} are 
quite similar to each other. In fact, these three patterns 
of magnetic fluxes give the same flavor structures, that is, 
the exactly same zero-mode contents (including the exotic 
modes and open string moduli ) survive 
after the orbifold projection. 
%( Strictly speaking, 
%there are just slightly differences with respect to the moduli dependence 
%of global factors of Yukawa matrices and K\"ahler metrics). 

As shown above, the orbifold projection is very useful to 
remove the extra modes. 
Furthermore, considering orbifold projections on the first and second
tori (or the first and third tori), 
there are some possibilities to obtain three generation structures with 
magnetic fluxes other than we show above and we study such possibilities
in the next section. 

%Finally, we have some comments. 
%In all the three, we obtain 
%the models with three generations of the quark 
%and lepton multiplets and six generations of the higgs multiplets. 
%Six generations of the higgs multiplets make it difficult to analyze 
%the models and obtain realistic ones. Then the simplest prescription is 
%that one assume a mechanism which make the five linear combinations of the 
%six heavy in the GUT scale(In Ref.~\cite{Abe:2012fj}, we have done that). 
%Then the five are decoupled from the light modes and the last one
%linear combination of the six remaining light is identified as the
%MSSM higgs fields. 

%Generally most of exotic fields are eliminated due to the effect of 
%chirality projection caused by magnetic fluxes. 
%However certain massless exotic modes and open string moduli 
%(massless diagonal components of superfields $\phi_i$) all of which feel zero f%luxes survive in our any case. 
%These exotics are severely constrained by many experimental 
%data at low energies. For the type1, we have pointed out the fact 
%in Ref.~\cite{Abe:2012fj} that all the exotic modes and 
%most of the open string moduli can be eliminated owing to 
%a certain orbifold projection on $r=2,3$ tori. 

One can generalize above results for N generations 
of the chiral matter fields with the replacing $``3\rightarrow {\rm N}"$. 
We strongly owe this generalization to the constraint that we 
have to derive generation structures on only the first torus to obtain 
full-rank Yukawa matrices. 

\subsection{$Z_2$ projection}

%Three generations of quarks and leptons are generated in 
%the first torus $r=1$ by the magnetic fluxes~(\ref{eq:model}). 
We consider the $Z_2$ projection, and then either even or odd modes of zero-modes 
remain. Zero-mode wavefunctions have the following relation~\cite{Abe:2008fi}, 
\begin{eqnarray*}
\phi^I(-z)=\phi^{M-I}(z).
\end{eqnarray*}
Using that, even and odd functions are given by,
\begin{gather}
\phi_{\rm
  even}^I(z)=\frac1{\sqrt2}\left(\phi^I(z)+\phi^{M-I}(z)\right) ,\nonumber\\
\phi_{\rm odd}^I(z)=\frac1{\sqrt2}\left(\phi^I(z)-\phi^{M-I}(z)\right).
\label{eq:evod}
\end{gather}
The degeneracy of zero-modes after the $Z_2$ projection is changed, and we show 
that in Table~\ref{tb:numzero}~\cite{Abe:2008fi}, 
\begin{table}[htb]
\begin{center}
\begin{tabular}{cccccccccccc} \hline
   $M$&0&1&2&3&4&5&6&7&8&9&10\\ \hline\hline
even&1&1&2&2&3&3&4&4&5&5&6\\
 odd&0&0&0&1&1&2&2&3&3&4&4
\\ \hline
\end{tabular}
\end{center}
\caption{The numbers of zero-modes for even and odd wavefunctions.}
\label{tb:numzero}
\end{table}
where $M=M_{ab}^{(r)}$ on each $r$-th torus.

One way to construct the plausible models by using the $T^6/Z_2$ orbifold 
is that we start with magnetic fluxes shown in Table~\ref{tb:3models} 
and then we assume the $T^6/Z_2$ orbifold where the $Z_2$ acts on the 
second and the third tori $r=2, 3$.
In this model building, we can eliminate extra modes 
without affecting three generations of quarks and leptons, 
as shown in Appendix A and Ref.~\cite{Abe:2012fj}.

On the other hand, from Table~\ref{tb:numzero}, 
we can see that there are other possibilities 
to obtain three generation structures by assuming that the $Z_2$ acts on the 
first and the second tori $r=1, 2$, or the first and third tori $r=1, 3$ : 
$M=4, 5$ for even modes and $M=7, 8$ for odd modes. 
Here we study such possibilities.
We concentrate on the case that the $Z_2$ projection acts 
on the first and the second tori $r=1, 2$ assuming that 
three generations are originated from the first torus, $r=1$. 
That is, the degeneracy numbers of quark and lepton zero-modes 
are equal to three for $r=1$ after $Z_2$ orbifolding, 
while the degeneracy numbers are equal to one for $r=2$ 
after orbifolding and one for the third torus. 
% (we will 
%summarize the results including the other case later). 
Then, $Z_2$ boundary conditions of 10D superfields 
$V$ and $\phi_i$ are assigned  as 
\begin{eqnarray}
V(x,-y_m,y_n) &=& +P V(x,y_m,y_n) P^{-1}, 
\nonumber \\
\phi_1(x,-y_m,y_n) &=& -P \phi_1(x,y_m,y_n) P^{-1}, 
\nonumber \\
\phi_2(x,-y_m,y_n) &=& -P \phi_2(x,y_m,y_n) P^{-1}, 
\nonumber \\
\phi_3(x,-y_m,y_n) &=& +P \phi_3(x,y_m,y_n) P^{-1}, 
\label{eq:z2tr} 
\end{eqnarray} for 
$^\forall m=4,5,6,7$ and $^\forall n=8,9$, 
where $P$ is a projection operator acting on YM indices 
satisfying $P^2={\bm 1}_N$. 
The $\phi_1$ and $\phi_2$ fields have the minus sign under the 
$Z_2$ reflection, because these are the vector fields, 
$A_i$ ($i=1,2$) on the $Z_2$ orbifold plane.
Note that the orbifold projection~(\ref{eq:z2tr}) respects 
the ${\cal N}=1$ supersymmetry preserved by the magnetic 
fluxes~(\ref{tb:3models}), because the $Z_2$-parities are 
assigned to the ${\cal N}=1$ superfields $V$ and $\phi_i$.

Choosing a proper projection operator $P$, we can construct 
the Pati-Salam models Eq.~(\ref{eq:pati}) with three generation 
chiral matter fields other than we showed 
in Table \ref{tb:3models}. However, 
on account of the orbifold projection~(\ref{eq:z2tr}), 
nonvanishing (continuous) Wilson-line parameters Eq.~(\ref{eq:wlm}) 
are possible
\footnote{Nonvanishing Wilson-line parameters 
would be possible also in the first and the second tori, 
if we allow non-zero VEVs of vector fields that are 
constants in the bulk but change their sign across the 
fixed points (planes) of the orbifold, that is beyond the 
scope of this paper. In this case localized magnetic fluxes 
at the fixed points might be induced which cause nontrivial 
effects on the wavefunction profile of the charged matter 
fields~\cite{Lee:2003mc}.} 
only on the third torus $r=3$. The Pati-Salam gauge group 
can be broken down as shown previously by the Wilson lines 
 Eq.~(\ref{eq:wlm}) on the third torus.
 Recall that the three generation structures are originated on the first 
torus.
However, in the above type of three-block magnetic fluxes, 
all the Yukawa matrices for 
the up-sector and down-sector of quarks, neutrinos, 
and charged leptons, have exactly the same form except a universal factor. 
The experimental values of their masses and mixings 
cannot be realized. Thus, we need 
consider the following five-block magnetic fluxes 
\begin{eqnarray}
M^{(i)} &=& \left( 
\begin{array}{ccccc}
M^{(i)}_C {\bm 1}_3 & \ & \ & \ & \ \\
\ & M^{(i)}_{C'} & \ & \ & \ \\
\ & \ & M^{(i)}_L {\bm 1}_2 & \ & \ \\
\ & \ & \ & M^{(i)}_{R'} & \ \\
\ & \ & \ & \ & M^{(i)}_{R''} 
\end{array} \right), 
\label{eq:mf5b}
\end{eqnarray}
where all the nonvanishing entries take values 
different from each other on at least the first tori. 
We search such a possibility 
in a systematic way. Then, it is found that it is impossible to construct 
three generation models with orbifold projections and five-block fluxes
(\ref{eq:mf5b}), 
since the SUSY preserving conditions become severer than the case 
with three-block fluxes. 
Obviously, that is the same in the case that the $Z_2$ acts 
on the first and the third tori $r=1, 3$. 

So far, we can obtain a unique 
flavor structure, which was studied in Ref.~\cite{Abe:2012fj}. 
We have considered only the simple fluxes like 
Eq.~(\ref{eq:ymvevs}). However, nonfactorizable fluxes which are magnetic 
fluxes crossing over two tori are studied in Ref.~\cite{Cremades:2004wa,Antoniadis:2009bg,DeAngelis:2012jc}, and it may enable 
us to obtain more various flavor structures and we study that 
in the next section.

\section{Nonfactorizable fluxes}
\label{sec:obl}
In this section, we study nonfactorizable flux models, 
where magnetic fluxes cross over two tori. 
We consider the case that the YM fields $A_i$ take the following VEVs 
instead of Eq.~(\ref{eq:ymvevs})
\begin{eqnarray}
&\vev{A_i} = \dfrac{\pi}{{\rm Im}\, \tau_i} 
M^{(i)}\,\bar{z}_{\bar{i}}+\dfrac{\pi}{{\rm Im}\, \tau_j} 
M^{(ij)}\,\bar{z}_{\bar{j}}+\bar\zeta_i ,\nonumber \\
&\vev{A_\mu} \ = \ 
\vev{\lambda_0} \ = \ 
\vev{\lambda_i} \ = \ 
\vev{F_i} \ = \ 
\vev{D} \ = \ 0, 
\label{eq:oblymvevs}
\end{eqnarray}
where $i\neq j$. This is a straightforward extension of Eq.~(\ref{eq:ymvevs}) 
and the additional second term of the VEV of $A_i$ corresponds 
to (Abelian) nonfactorizable fluxes, 
\begin{eqnarray}
M^{(ij)} &=&\left( 
\begin{array}{cccc}
M^{(ij)}_1 {\bm 1}_{N_1} & \ & \ & \ \\
\ & M^{(ij)}_2 {\bm 1}_{N_2} & \ & \ \\
\ & \ & \ddots & \ \\
\ & \ & \ & M^{(ij)}_n {\bm 1}_{N_n} \ \\
\end{array} \right). \label{eq:oblflux} \\
\nonumber
\end{eqnarray}
The VEV Eq.~(\ref{eq:oblymvevs}) trivially satisfies the F term SUSY condition 
Eq.~(\ref{eq:abfi}) as well as the VEV Eq.~(\ref{eq:ymvevs}) does. 
That is because the complex structure in the second term of the first
line (\ref{eq:oblymvevs}) is consistent with the one in the first
term.
If we take a different coefficient in the second term, the SUSY
condition is not satisfied.
%However, note that F-term SUSY condition Eq.~(\ref{eq:abfi}) constraint how 
%to introduce nonfactorizable fluxes. If we consider a VEV which have 
%other forms (e.g. $M{\rm Im}\,z$, $M'{\rm Re}\,\bar z$ and etc.), 
%the condition strongly works. In contrast, we consider the VEVs of 
%complex scalar fields $A_i$ respecting complex structures, so each term of 
%Eq.~(\ref{eq:oblymvevs}) satisfy a certain relation in terms of 
%real coordinates and real component fields automatically. 
For magnetic fluxes crossing over the two tori, e.g. the first and second tori, 
there are four elements, $F_{x_1,x_2}, F_{x_1,y_2}, 
F_{y_1,x_2}$, and $F_{y_1,y_2}$ in the real basis, where 
 $z_1=\frac12(x_1+\tau_1y_1)$ for the first torus and 
$z_2=\frac12(x_2+\tau_2y_2)$ for the second torus.
In order to derive a well defined wavefunction, 
only $F_{x_1,y_2}$ and $F_{y_1,x_2}$ are allowed with a certain SUSY 
relation \cite{Cremades:2004wa,Antoniadis:2009bg}. 
The VEV Eq.~(\ref{eq:oblymvevs}) 
corresponds to such a case exactly.

With the above nonfactorizable fluxes, the numbers of degenerate zero-modes 
$(V^{n=0})_{ab}$ and $(\phi^{n=0}_j)_{ab}$ are changed. 
$(V^{n=0})_{ab}$ obtain mass terms with gauge symmetry breaking, so 
we explain about chiral superfields $\phi_j$. 
For the moment, we consider only the case $i,j=1, 2$ and  
$M^{(12)}$ and $M^{(21)}$ cross over between the first and the second 
tori.
Later we will discuss its extensions, but such extensions do not 
lead to interesting models.

Now, we define the matrix
%%%modified M -> M^T (130418)
\begin{eqnarray}
\mathcal M_{ab} &=& 
\left(\begin{array}{ccc}
m^{(1)}_{ab} & m^{(12)}_{ab} \\
m^{(21)}_{ab} & m^{(2)}_{ab} 
\end{array}\right)\nonumber \\
m^{(i)}_{ab} &=& M^{(i)}_{a}-M^{(i)}_{b}
\nonumber \\
m^{(ij)}_{ab} &=& m^{(ij)}_{a}-m^{(ij)}_{b}
\nonumber \\
m^{(ij)}_{a} &=& \frac{{\rm Im}\, \tau_i}
{{\rm Im}\, \tau_j}M_a^{(ij)}+M_a^{(ji)}.
\label{eq:obsimple}
\end{eqnarray}
Diagonal elements $m^{(i)}_{ab}$ correspond to the 
fluxes introduced in 
Sec.~\ref{sec:mag}, and off-diagonal ones are nonfactorizable fluxes. 
We have to impose Riemann conditions in order to have a 
well-defined zero-mode wavefunction, and then the matrix $\mathcal M$ 
and complex  
structure $\tau_i$ have to satisfy the following constraints 
\cite{Cremades:2004wa}
\begin{eqnarray}
\mathcal M^{ij}\in \bm Z,\qquad (\mathcal M\cdot{\rm Im}\, \Omega)^T = 
\mathcal M\cdot{\rm Im}\,\Omega,\qquad \mathcal M\cdot{\rm Im}\, \Omega > 0,
\label{eq:riemanncond}
\end{eqnarray} 
where $\Omega \equiv {\rm diag}\, (\tau_1, \tau_2)$ is a 
general complex structure 
\footnote{In this paper, we unalterably focus on factorizable torus. We 
can consider generalized torus with off-diagonal element of 
$\Omega$ \cite{Cremades:2004wa, Antoniadis:2009bg, DeAngelis:2012jc}, 
however, it is difficult to study 
that in a systematic way. }. 
Then, $\det\mathcal M$ determines the number of 
degenerate zero-modes on the two tori, and the total degeneracy is equal to 
$|\det\mathcal M\times M^{(3)}_{ab}|$. 
While the zero-mode wavefunctions on the third torus are obtained by 
Eq.~(\ref{eq:zerowave}) still, the zero-mode wavefunctions on 
the other two tori are written by the following forms 
\footnote{ Here, we take another gauge according to Ref.~\cite{Antoniadis:2009bg} : $A_i\rightarrow A_i-\partial_i\chi^{(i)}$, where 
$\partial_1\chi^{(1)}=\frac1{\mathrm{Im}\,\tau_2}M^{(12)}\mathrm{Re}\,z_2$ and 
$\partial_2\chi^{(2)}=\frac1{\mathrm{Im}\,\tau_1}M^{(21)}\mathrm{Re}\,z_1$ }, 
\begin{gather}
\phi^{\vec{j}_{ab}, \mathcal{M}_{ab}} (\vec{z}, \Omega) = \mathcal{N} \cdot e^{i \pi [\mathcal{M}_{ab}\cdot \vec{z}] \cdot (\mathcal{M}_{ab} \cdot \mathrm{Im}\,\Omega)^{-1} \mathrm{Im}\,[\mathcal{M}_{ab}\cdot \vec{z}] } \cdot \vartheta \begin{bmatrix}\vec{j}_{ab}\\[5pt] 0\end{bmatrix} (\mathcal{M}_{ab}\cdot \vec{z}, \mathcal{M}_{ab}\cdot \Omega),\label{eq:zerowave2}
\end{gather}
where $\vec z=\frac{1}{2}(\vec x + \Omega\vec{y})$,  $\vec j_{ab}$ labels the degeneracy generated on the two tori, 
and $\vartheta$ is the Riemann theta-function, which is a generalization 
of the Jacobi theta-function Eq.~(\ref{eq:jacobi}), 
\begin{gather}
\vartheta\begin{bmatrix}\vec{a}\\[5pt] \vec{b}\end{bmatrix} (\vec{\nu}, \Omega) = \sum_{\vec{m} \in \mathbb{Z}^n} e^{\pi(\vec{m}+\vec{a}) \cdot \Omega \cdot (\vec{m}+\vec{a})} e^{2 \pi i (\vec{m}+\vec{a}) \cdot (\vec{\nu}+\vec{b})}.
\label{eq:riemann}
\end{gather}
Normalization is defined in a way similar to the previous one. 
For more details, see Ref.\cite{Cremades:2004wa,Antoniadis:2009bg}.

The constraints Eq.~(\ref{eq:riemanncond}) and the wavefunction 
Eq.~(\ref{eq:zerowave2}) are valid for a field $\phi_3$ which have (totally) 
positive chirality  
on the two tori. 
%Here, we choose ${\rm Im} \tau_i$ to be positive (that is equivalent to choosing fundamental region. 
As for $\phi_1$ and $\phi_2$ which have the negative chirality 
$(+, -)$ or $(-, +)$ on the two tori, they need to be mixed up to get a 
zero-mode wavefunctions. Then, according to Ref.~\cite{Antoniadis:2009bg}, 
we parametrize them as following,
\begin{eqnarray}
\phi_1^{ab}=\alpha^{ab} \Phi^{ab}\hspace{5em}\phi_2^{ab}=\beta^{ab} \Phi^{ab}.
\label{eq:mixedup}
\end{eqnarray}
The condition to obtain a well-defined wavefunction Eq.~(\ref{eq:riemanncond}) 
and an explicit form of the wavefunction Eq.~(\ref{eq:zerowave2}) can also 
be applied for $\Phi$ with replacing $\Omega$ 
by $\tilde\Omega_{ab} = \hat\Omega_{ab}\cdot\Omega$, where
\begin{eqnarray*}
\hat\Omega_{ab}=\frac1{1+q_{ab}^2}\left(\begin{array}{cc}
1-q_{ab}^2 & -2q_{ab}\\
-2q_{ab} & q_{ab}^2-1
\end{array}\right). 
\end{eqnarray*}
Mixing parameters $q_{ab}=\beta_{ab}/\alpha_{ab}$ are given for individual  
bifundamental representations and their values are determined by the second one of Eq.~(\ref{eq:riemanncond}). 
Since $\det\hat\Omega =-1$, 
one can see ${\rm sign}\,(\det\mathcal M)$ relates to the chirality of 
a field from the last one of Eq.~(\ref{eq:riemanncond}).

\subsection{Nonfactorizable flux Ansatz}
We study configurations of nonfactorizable magnetic fluxes. 
Similarly to the previous ones, we aim to realize a zero-mode spectrum in 
10D magnetized $U(8)$ SYM theory, that contains the MSSM with the 
gauge symmetry $SU(3)_C\times SU(2)_L\times U(1)_Y$ and three generations of 
the quark and the lepton chiral multiplets.

If there are only  $M^{(ij)}$ fluxes with $M^{(i)}=0$, we obtain 
the same result as the factorizable models shown in Sec.~\ref{sec:mag} 
with changing definition of 
torus cycles, so we consider both of non-vanishing 
$M^{(i)}$ and $M^{(ij)}$ simultaneously. 
In the previous section, we show the simplest case 
with nonfactorizable fluxes (\ref{eq:oblymvevs}). 
Since we compactify the 10D SYM theory on three tori, 
one can think the magnetic flux background, where 
the VEVs of $A_i$ in Eq.~(\ref{eq:oblflux}) have 
another term, $\frac{\pi}{{\rm Im}\, \tau_k} 
M^{(ik)}\,\bar{z}_{\bar{k}}$ for $k\neq i,j$. However, 
such general cases 
%won't give us three generation models 
%due to 
have strong constraints to realize a 
well-defined wavefunction, that is Eq.~(\ref{eq:riemann}). 
At any rate, we search such possibilities in a systematic way with 
all the flux parameters $M_a^{(i)}$ and $m_{a}^{(ij)}$ being in the range 
between $-10$ and 10, but we could not obtain three generation models 
in the parameter range. 
Thus, it is adequate for us to concentrate on the magnetic fluxes 
crossing over the first and second tori, Eqs.~(\ref{eq:oblymvevs}) and 
(\ref{eq:obsimple}).
That is, we consider the cases, where non-vanishing magnetic fluxes 
correspond to ${M^{(12)}}$, ${M^{(21)}}$ and ${M^{(i)}}$ ($i=1,2,3$) and 
the others are vanishing. 
%This is why we sad it's adequate for us to show the simplest case. 
%We consider cases with factorizable fluxes 
%on the three tori and nonfactorizable fluxes on only the two tori, 
%and we choose the first and the second tori as that (Although 
%alternative choice can be possible, of course, it will not lead 
%new landscape other than the one ). 
In the following, 
we search possible flux configurations in a systematic way, 
where the three generation structure is 
generated on the first and the second tori by the nonfactorizable fluxes.

With the above setup, the three-block flux (Pati-Salam) model is 
the simplest. Two-block models like Eq.~(\ref{eq:abelianflux}) 
are improbable in a way similar to the case with only the factorizable fluxes. 
%We try to construct such three-block models and can easily see that 
%there is an infinite number 
%of possibilities to have three generation models, because there is an 
%infinite number of $2\times2$ matrices whose determinant is equal to three. 
In the three-block models, 
we need the Wilson lines to break the Pati-Salam gauge group 
as well as in Sec.~\ref{sec:mag}. If we could construct four-block or 
five-block models,  
e.g., like Eq.~(\ref{eq:mf5b}), we would not necessarily need 
the Wilson lines. We also search a possibility 
to give such an attractive model in the systematic way 
with all the flux parameters $M_a^{(i)}$ and $m_{ab}^{(ij)}$ being in the range between $-10$
and 10, but 
we could not find a four-block model or a five-block model, 
because of the severer constraints for preserving SUSY and having 
well-defined wavefunctions than the Pati-Salam models . 
Thus, only the three-block Pati-Salam models are possible like 
the models with only the factorizable fluxes in Sec.~\ref{sec:mag}. 
In the next section, we show the details of our systematic research about 
the Pati-Salam models and its results.

\subsection{Pati-Salam model}

In this section, we concentrate on the three-block flux models with 
three (Abelian) factorizable fluxes $M^{(i)}~(i=1,2,3)$ 
and (Abelian) nonfactorizable fluxes $M^{(12)}$ and $M^{(21)}$. 
%Since it is difficult to deal with varying these five (matrix) 
%parameters due to the constraints 
%Eq.~(\ref{eq:obsimple}) and (\ref{eq:riemann}), 
We use $m^{(12)}_{a}$ and $m^{(21)}_{a}$ as parameters 
instead of $M^{(12)}_{a}$ and $M^{(21)}_{a}$, 
such as they satisfy the following relation 
\begin{eqnarray*}
{\rm Im}\, \tau_1m^{(21)}_{a} = 
{\rm Im}\, \tau_2m^{(12)}_{a},
\end{eqnarray*}
that is nothing but the SUSY preserving condition 
and satisfied automatically in Eq.~(\ref{eq:obsimple}). 
There are an infinite number of possibilities,  
so we restrict the values of the parameters to the range 
between $-10$ and $10$. As a result, we find many possibilities for three generation models. 

There are two types of models. 
The one, which we call ``type 1", 
is that 
Higgs multiplets come from $\phi_3$ which have the 
chirality $(-, -, +)$. In this case left-handed and right-handed matter 
fields come from $\phi_1$ and $\phi_2$ both of which have the negative 
chirality totally on the first two tori. 
In the other case, ``type2", either left-handed or right-handed matter 
fields come from $\phi_3$, and then 
they have the opposite chirality totally on the two tori. 

\begin{figure}[htbp]
\begin{center}
\includegraphics[width=0.6\textwidth, clip]{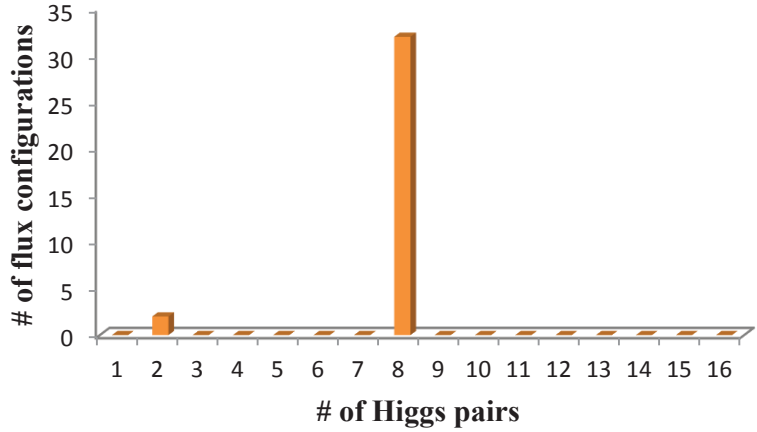}\\
\includegraphics[width=0.6\textwidth, clip]{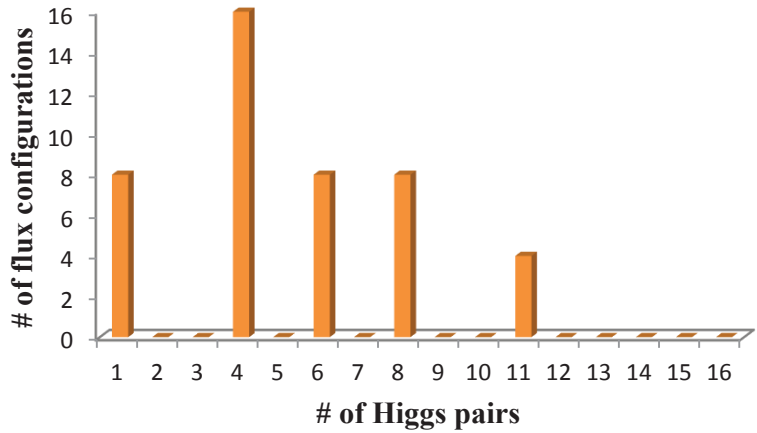}
\end{center}
 \caption{The number of magnetic flux configurations for each number
   of Higgs pairs 
in the 
type1 models where Higgs multiplets come from $\phi_3$. The upper
panel corresponds to the case with $M^{(3)}={\rm diag}\,(0, 1, -1)$ 
%which doubles the number of Higgs multiplets. 
and the lower panel corresponds to  
$M^{(3)}={\rm diag}\,(0, 1, 0)$ or ${\rm diag}\,(0, 0, -1)$ %doesn't have an 
%affect on the number of Higgs multiplets
.}
 \label{fig:nh12}
\end{figure}

First we show the result about the type1 models in Fig.~\ref{fig:nh12}. 
In this case, $M^{(3)}$ have the three possibilities, 
$M^{(3)}={\rm diag}\,(0, 1, -1)$, ${\rm diag}\,(0, 1, 0)$ and
 ${\rm diag}\,(0, 0, -1)$ without affecting the 
three generation structures of quark and lepton multiplets generated 
on the first and the second tori. 
The upper panel of Fig.~\ref{fig:nh12} corresponds to the case with 
$M^{(3)}={\rm diag}\,(0, 1, -1)$.
The number of Higgs pairs is 
increased twofold by $M^{(3)}$ on the third torus, and 
all the numbers of Higgs pairs are even. 
The lower panel shows the case that $M^{(3)}={\rm diag}\,(0, 1, 0)$  
and ${\rm diag}\,(0, 0, -1)$.

We see that only particular numbers 
of Higgs pairs are allowed from the two panels. 
In previous section, our three generation 
models inevitably include six pairs of Higgs multiplets. 
Now from the lower panel, 
we find the possibilities to obtain the models with one pair of Higgs 
doublets just like  the MSSM. It is quite attractive and we show 
the details in the next section.

\begin{figure}[htbp]
\begin{center}
\includegraphics[width=0.6\textwidth, clip]{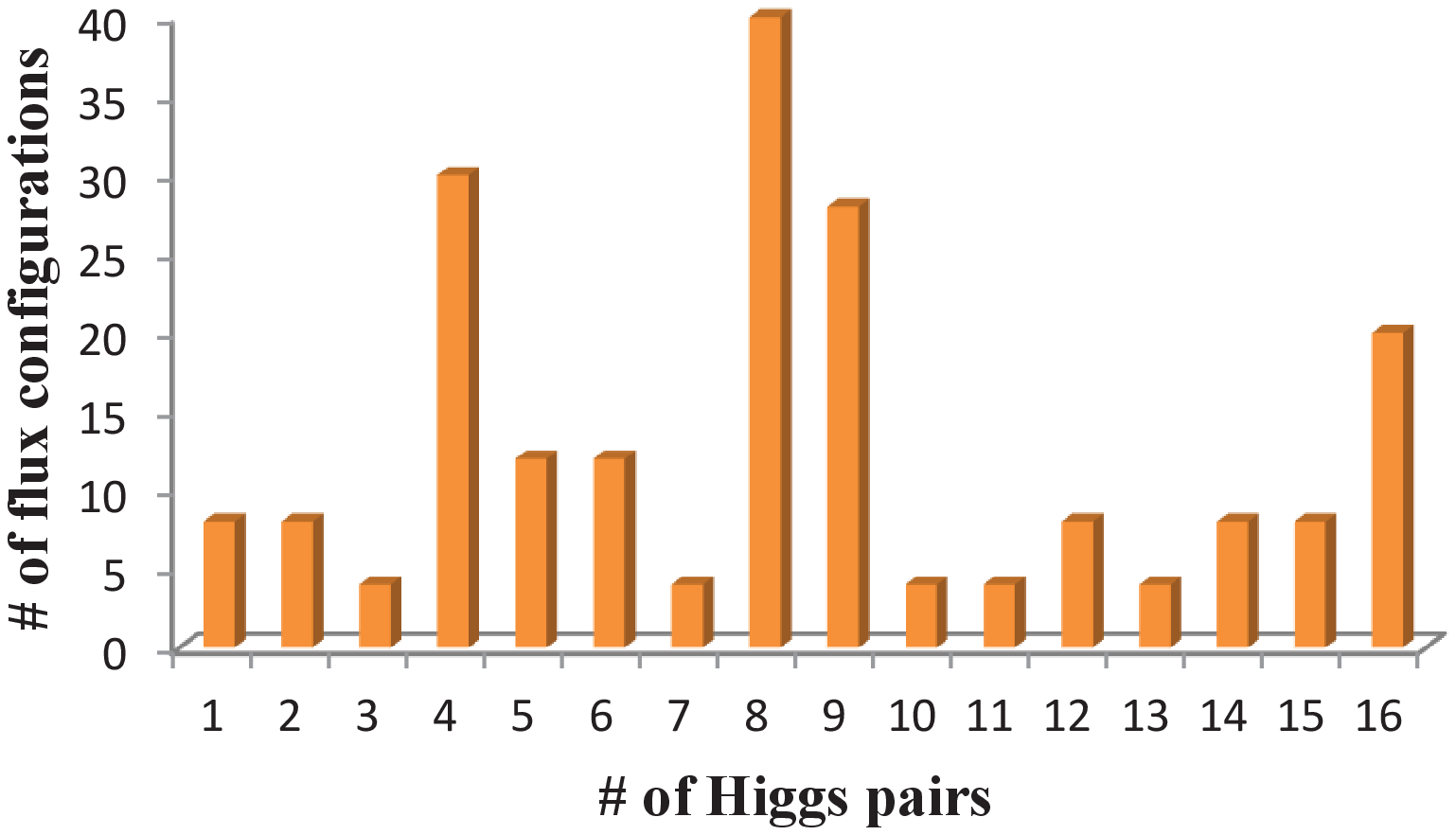}
\end{center}
 \caption{The number of magnetic fluxes for each number of Higgs pairs  in the 
type2 models where Higgs multiplets come from the mixture of $\phi_1$ and $\phi_2$.}
 \label{fig:ob13}
\end{figure}

Next, in Fig.~\ref{fig:ob13} we show the result about the type2 models 
where left-handed and right-handed matter fields have the opposite chirality 
on the first two tori\footnote{
We concentrate only the case in which the right-handed matter fields come 
from $\phi_3$ .}. What is quite different from the type1 is that all the 
numbers of Higgs pairs up to 16 are allowed  (at least in our searching region). 
One-pair Higgs models are also possible. 

We find many possibilities to construct the models consistent with 
the MSSM and we obtain more various flavor structures.
In particular, nonfactorizable fluxes enable us to construct 
one-pair Higgs  models. 
We show the details of such typical models in the following. 

\subsection{One-pair Higgs models}
\label{sec:s2hd}
Here, we focus on the models, 
which have a pair of  Higgs doublets just like the MSSM. 
%From now on, we call that one-pair Higgs model. 
There are various flux configurations to give such models. 
Now, let us show details of a typical one included 
in the type1 models. 
%in which the left-handed 
%and right-handed matter fields have the same chirality totally 
%on the two tori and that is included in the lower panel of 
%Fig.~\ref{fig:nh12}. 
The flux configuration is set as, 
\begin{align*}
(M^{(1)}_C, M^{(1)}_L, M^{(1)}_R) ~~ &= (0,-2,-1),\\
(M^{(2)}_C, M^{(2)}_L, M^{(2)}_R) ~~ &= (0,+1,+2),\\
(M^{(3)}_C, M^{(3)}_L, M^{(3)}_R) ~~ &= (0,0,-1),\\
(M^{(12)}_C, M^{(12)}_L, M^{(12)}_R) &= (0,+1,+1),\\
(M^{(21)}_C, M^{(21)}_L, M^{(21)}_R) &= (0,+1,+1).\\
\end{align*}
Then, the differences of fluxes which appear in zero-mode 
equations of multiplets are obtained as
\begin{gather*}
\mathcal M_{L} = \left(\begin{array}{cc}+2&-1\\ -1&-1\end{array}\right),
\qquad \mathcal M_{R} = \left(\begin{array}{cc}-1&+1\\ +1&+2\end{array}\right),
\qquad \mathcal M_{H}= \left(\begin{array}{cc}-1&0\\ 0&-1\end{array}\right),\\
M^{(3)}_C -M^{(3)}_L =0,
\quad\qquad M^{(3)}_R -M^{(3)}_C =-1,
\quad\qquad M^{(3)}_L -M^{(3)}_R=+1,
\end{gather*}
where a subscript $L$ in $\mathcal M_L$ means left-handed matter fields, 
$R$ does right-handed matter and $H$ does Higgs fields. 
The absolute values of the  determinants, $\mathcal M_L$ and $\mathcal M_R$, are equal to
three and they originate three generation structures. 
In contrast, a determinant of $\mathcal M_H$ is equal to one, and 
fluxes on the third torus do not have any effects on the number 
of degenerate zero-modes. 
With these fluxes, we have the following zero-mode contents, 
block off-diagonal parts of $\phi_1$ and $\phi_2$, which are 
mixed up shown in Eq.~(\ref{eq:mixedup})\footnote{
In the model, Higgs sectors does not feel the nonfactorizable fluxes 
($m^{(ij)}_L-m^{(ij)}_R=0$), so 
$\phi_1$ and $\phi_2$ are not mixed up in the parts. 
However, they all are eliminated 
by the factorizable fluxes on the third torus $M^{(3)}$. 
We simply show the result 
like this. }: 
\begin{align*}
\Phi^{{\cal I}_{ab}} & =
\left( 
\begin{array}{cc|c|cc}
& & Q^I & 0 & 0 \\
& & L^I & 0 & 0 \\
\hline 
0 & 0 & & 0 & 0 \\
\hline 
U^J & N^J & 0 & & \\
D^J & E^J & 0 & & 
\end{array}
\right), 
\nonumber 
\end{align*}
block diagonal parts of $\phi_1$ and $\phi_2$, which do not feel the fluxes 
and each of zero modes exists separately: 
\begin{align*}
\phi_1^{{\cal I}_{ab}} & =
\left( 
\begin{array}{cc|c|cc}
\Omega_C^{(1)} & \Xi_{CC'}^{(1)} & & & \\
\Xi_{C'C}^{(1)} & \Omega_{C'}^{(1)} & & & \\
\hline 
 & & \Omega_L^{(1)} & & \\
\hline 
 & & & \Omega_{R'}^{(1)} & \Xi_{R'R''}^{(1)} \\
 & & & \Xi_{R''R'}^{(1)} & \Omega_{R''}^{(1)} 
\end{array}
\right),~~
\phi_2^{{\cal I}_{ab}} = 
\left( 
\begin{array}{cc|c|cc}
\Omega_C^{(2)} & \Xi_{CC'}^{(2)} & & & \label{phis1} \\
\Xi_{C'C}^{(2)} & \Omega_{C'}^{(2)} & & & \\
\hline 
 & & \Omega_L^{(2)} & & \\
\hline 
 & & & \Omega_{R'}^{(2)} & \Xi_{R'R''}^{(2)} \\
 & & & \Xi_{R''R'}^{(2)} & \Omega_{R''}^{(2)} 
\end{array}
\right),
\end{align*}
and the field $\phi_3$:
\begin{align*}
\phi_3^{{\cal I}_{ab}} &= 
\left( 
\begin{array}{cc|c|cc}
\Omega_C^{(3)} & \Xi_{CC'}^{(3)} & 0 & 
0 & 0 \\
\Xi_{C'C}^{(3)} & \Omega_{C'}^{(3)} & 0 & 
0 & 0 \\ 
\hline 
0&0  & \Omega_L^{(3)} & 
H_u & H_d \\ 
\hline 
0 & 0 & 0 & \Omega_{R'}^{(3)} & \Xi_{R'R''}^{(3)} \\
0 & 0 & 0 & \Xi_{R''R'}^{(3)} & \Omega_{R''}^{(3)} 
\end{array}
\right),
\end{align*}
where we use the same notations as we use in the previous section, 
however, Higgs doublets do not have a generation index. 
The three generation chiral matter fields and one pair of Higgs fields 
are realized just like the MSSM. Although all the chiral exotic fields 
are eliminated by the fluxes, there are some vector-like exotic modes 
and open string moduli. 
With all flux configurations which give one-pair Higgs type1 models, the same contents are generated. 

We show the explicit form of Yukawa matrices of this model in 
Appendix \ref{sec:appb}. According to that, 
Yukawa matrices can be full-rank and hierarchical, and 
all the elements of Yukawa matrices 
can take nonzero values different from each other. 
Thus, it would be  possible to give 
a realistic pattern of quark and lepton masses and their mixings. 
We will study numerically these values elsewhere.

In Table~\ref{tb:s2hd12} we show all the patterns of magnetic flux
configurations, which give such one-pair Higgs models 
included in the type1.
We choose $M_C=0$ 
and the first one (No. 1) corresponds to the above.
\begin{table}
\begin{center}
\begin{tabular}{|c|cccccc|cccc|} \hline
& $M_L^{(1)}$& $M_R^{(1)}$& $M_L^{(2)}$& $M_R^{(2)}$& $M_L^{(3)}$& $M_R^{(3)}$& $m_L^{(12)}$& $m_R^{(12)}$& $m_L^{(21)}$& $m_R^{(21)}$\\ \hline\hline
No. 1& $-2$& $-1$& $+1$& $+2$& $0$& $-1$& $+1$& $+1$& $+1$& $+1$\\ \hline
No. 2& $-2$& $-1$& $+1$& $+2$& $+1$& $0$& $+1$& $+1$& $+1$& $+1$\\ \hline
No. 3& $+1$& $+2$& $-2$& $-1$& $0$& $-1$& $+1$& $+1$& $+1$& $+1$\\ \hline
No. 4& $+1$& $+2$& $-2$& $-1$& $+1$& $0$& $+1$& $+1$& $+1$& $+1$\\ \hline
No. 5& $-2$& $+3$& $+1$& $+2$& $0$& $-1$& $+1$& $+3$& $+1$& $+3$\\ \hline
No. 6& $+1$& $+2$& $-2$& $+3$& $0$& $-1$& $+1$& $+3$& $+1$& $+3$\\ \hline
No. 7& $-3$& $+2$& $-2$& $-1$& $+1$& $0$& $+3$& $+1$& $+3$& $+1$\\ \hline
No. 8& $-2$& $-1$& $-3$& $+2$& $+1$& $0$& $+3$& $+1$& $+3$& $+1$\\ \hline
\end{tabular}
\end{center}
\caption{One-pair Higgs models in type1.}
\label{tb:s2hd12}
\end{table}

Next, we study the details of one-pair Higgs models 
included in the type2 models.  
We show one example of flux configurations 
%and their zero-mode contents briefly using the same notations. 
%The flux configuration is obtained 
as follows, 
\begin{align*}
(M^{(1)}_C, M^{(1)}_L, M^{(1)}_R) ~~ &= (0,-1,-1)\\
(M^{(2)}_C, M^{(2)}_L, M^{(2)}_R) ~~ &= (0,+2,-7)\\
(M^{(3)}_C, M^{(3)}_L, M^{(3)}_R) ~~ &= (0,0,+1)\\
(M^{(12)}_C, M^{(12)}_L, M^{(12)}_R) &= (0,-1,-2)\\
(M^{(21)}_C, M^{(21)}_L, M^{(21)}_R) &= (0,-1,-2),\\
\end{align*}
then 
\begin{gather}
\mathcal{M}_L = \left(\begin{array}{cc}+1& +1\\ +1&-2\end{array}\right),
\qquad \mathcal{M}_R = \left(\begin{array}{cc}-1&-2\\ -2&-7\end{array}\right),
\qquad \mathcal{M}_H= \left(\begin{array}{cc}0&+1\\ +1&+9\end{array}\right),
\nonumber\\
M^{(3)}_C -M^{(3)}_L =0,
\quad\qquad M^{(3)}_L -M^{(3)}_R =+1,
\quad\qquad M^{(3)}_R -M^{(3)}_C=-1.
\nonumber
\end{gather}
The zero-mode contents are as follows, 
block off-diagonal parts of $\phi_1$ and $\phi_2$, which are 
mixed up shown in Eq.~(\ref{eq:mixedup}): 
\begin{align*}
\Phi^{{\cal I}_{ab}} & =
\left( 
\begin{array}{cc|c|cc}
& & Q^I & 0 & 0 \\
& & L^I & 0 & 0 \\
\hline 
0 & 0 & & H_u & H_d \\
\hline 
0 & 0 & 0 & & \\
0 & 0 & 0 & & 
\end{array}
\right), 
\nonumber 
\end{align*}
block diagonal parts of $\phi_1$ and $\phi_2$, which do not feel the fluxes 
and each of zero modes exists separately: 
\begin{align*}
\phi_1^{{\cal I}_{ab}} & =
\left( 
\begin{array}{cc|c|cc}
\Omega_C^{(1)} & \Xi_{CC'}^{(1)} & & & \\
\Xi_{C'C}^{(1)} & \Omega_{C'}^{(1)} & & & \\
\hline 
 & & \Omega_L^{(1)} & & \\
\hline 
 & & & \Omega_{R'}^{(1)} & \Xi_{R'R''}^{(1)} \\
 & & & \Xi_{R''R'}^{(1)} & \Omega_{R''}^{(1)} 
\end{array}
\right),~~
\phi_2^{{\cal I}_{ab}} = 
\left( 
\begin{array}{cc|c|cc}
\Omega_C^{(2)} & \Xi_{CC'}^{(2)} & & & \label{phis1} \\
\Xi_{C'C}^{(2)} & \Omega_{C'}^{(2)} & & & \\
\hline 
 & & \Omega_L^{(2)} & & \\
\hline 
 & & & \Omega_{R'}^{(2)} & \Xi_{R'R''}^{(2)} \\
 & & & \Xi_{R''R'}^{(2)} & \Omega_{R''}^{(2)} 
\end{array}
\right),
\end{align*}
and the field $\phi_3$:
\begin{align*}
\phi_3^{{\cal I}_{ab}} &= 
\left( 
\begin{array}{cc|c|cc}
\Omega_C^{(3)} & \Xi_{CC'}^{(3)} & 0 & 
0 & 0 \\
\Xi_{C'C}^{(3)} & \Omega_{C'}^{(3)} & 0 & 
0 & 0 \\ 
\hline 
0&0  & \Omega_L^{(3)} & 
0 & 0 \\ 
\hline 
U^J & N^J & 0 & \Omega_{R'}^{(3)} & \Xi_{R'R''}^{(3)} \\
D^J & E^J & 0 & \Xi_{R''R'}^{(3)} & \Omega_{R''}^{(3)} 
\end{array}
\right).
\end{align*}
In this model, zero-mode contents are the same as the previous ones in
type1. 
%We can also see the followings by 
%analytic study, 
Furthermore, Yukawa matrices can be full-rank and hierarchical, and 
all the elements of Yukawa matrices 
can take non-zero values different from each other. 

There are eight flux configurations 
which give one-pair Higgs models included in the type2 model, 
and these are shown in Table~\ref{tb:s2hd13}. 
The first one, No. 1, in Table~\ref{tb:s2hd13} corresponds to the above. 
The four patterns from No. 1 to No. 4 give the same zero-mode
contents. 
However, the zero-modes of  
the last four models, No. 5-8, include additional fields. 
In these models, the Higgs sector does not feel the nonfactorizable fluxes 
($m^{(ij)}_L-m^{(ij)}_R = 0$ ) and we get their zero-modes separately 
in $\phi_1$ and $\phi_2$. Furthermore, the factorizable fluxes on the third 
torus $M^{(3)}$ have no effect on the Higgs sector. 
That is the significant difference from models, No. 1-4 in Table~\ref{tb:s2hd12}. 
Then, the models have 
the MSSM Higgs correctly in $\phi_1$ and also have representations conjugate to the Higgs pair but can not couple with quarks and leptons in $\phi_2$. 

\begin{table}
\begin{center}
\begin{tabular}{|c|cccccc|cccc|} \hline
& $M_L^{(1)}$& $M_R^{(1)}$& $M_L^{(2)}$& $M_R^{(2)}$& $M_L^{(3)}$& $M_R^{(3)}$& $m_L^{(12)}$& $m_R^{(12)}$& $m_L^{(21)}$& $m_R^{(21)}$\\ \hline\hline
No. 1& $-1 $&$ -1 $&$ +2 $&$ -7 $&$ 0 $&$ +1 $&$ -1 $&$ -2 $&$ -1 $&$ -2$\\ \hline
No. 2& $2$&$ -7 $&$ -1 $&$ -1 $&$ 0 $&$ +1 $&$ -1 $&$ -2 $&$ -1 $&$ -2$\\ \hline
No. 3& $-1 $&$ -4 $&$ +2 $&$ -3 $&$ 0 $&$ +1 $&$ +1 $&$ -3 $&$ +1 $&$ -3$\\ \hline
No. 4& $2 $&$ -3 $&$ -1 $&$ -4 $&$ 0 $&$ +1 $&$ +1 $&$ -3 $&$ +1 $&$ -3$\\ \hline
No. 5& $-2 $&$ -3 $&$ -9 $&$ -8 $&$ +1 $&$ +1 $&$ +3 $&$ +3 $&$ +7 $&$ +7$\\ \hline
No. 6& $-2 $&$ -3 $&$ -9 $&$ -8 $&$ +1 $&$ +1 $&$ +7 $&$ +7 $&$ +3 $&$ +3$\\ \hline
No. 7& $-2 $&$ -3 $&$ -9 $&$ -8 $&$ +1 $&$ +1 $&$ -3 $&$ -3 $&$ -7 $&$ -7$\\ \hline
No. 8& $-2 $&$ -3 $&$ -9 $&$ -8 $&$ +1 $&$ +1 $&$ -7 $&$ -7 $&$ -3 $&$ -3$\\ \hline
\end{tabular}
\end{center}
\caption{One-pair Higgs models in  type2.}
\label{tb:s2hd13}
\end{table}

\subsection{Nonfactorizable flux models on $T^6/Z_2$ orbifolds}

There are other possibilities for model building.
We can consider nonfactorizable fluxes with 
orbifold projection, where magnetic fluxes have to be five-block forms because 
(continuous) Wilson lines cannot be introduced to break the 
gauge group by the orbifold projection. 
When we consider the $Z_2$ projection Eq.~(\ref{eq:z2tr}) 
either even or odd modes of zero-modes remain. 
How to derive even and odd functions 
from Eq.~(\ref{eq:zerowave}) and (\ref{eq:jacobi}) has already known 
for factorizable fluxes in 
Ref.\cite{Abe:2008fi}, and we find that even and odd functions can be 
derived from Eq.~(\ref{eq:zerowave2}) and (\ref{eq:riemann}) as natural 
generalization of the factorizable case Eq.~(\ref{eq:evod}) as,
\begin{align*}
\phi_\mathrm{even}^{\vec{j}}(\vec{z}) = \frac{1}{\sqrt{2}} (\phi^{\vec{j},{\mathcal M}}(\vec{z})+\phi^{\vec{e}-\vec{j},{\mathcal M}}(\vec{z})), \\
\phi_\mathrm{odd}^{\vec{j}}(\vec{z}) = \frac{1}{\sqrt{2}} (\phi^{\vec{j},{\mathcal M}}(\vec{z})-\phi^{\vec{e}-\vec{j},{\mathcal M}}(\vec{z})) ,
\end{align*}
where,
\begin{gather}
\vec{e} \equiv \vec{e}_1 + \vec{e}_2 = \mat{1\\ 0} + \mat{0\\ 1} =\mat{1\\ 1}.
\nonumber
\end{gather}
Note that we use the following relations to obtain these functions, 
\begin{gather}
\phi^{\vec{j},{\mathcal M}}(-\vec{z}\hspace{1pt})=\phi^{\vec{e}-\vec{j},{\mathcal M}}(\vec{z}\hspace{1pt}) \label{prop}.
\nonumber
\end{gather}

According to that, we find that a rule for the number of the degenerate zero-modes 
after a $Z_2$ projection is the same as factorizable cases and 
we see the rule in Table \ref{tb:numzero} with replacing 
$M$ by $\det\mathcal M$. \footnote{
 Strictly speaking, there are some exceptional 
case, however, it has nothing to do with us as far as we aim to the 
three generation models.}

On such a background, we search flux configurations to give 
the three generation models in a systematic way and the range
between $-10$ and 10. 
As a result, we could not construct three generation models with
four-block or five-block fluxes because SUSY conditions are severe.

\section{Conclusions and discussions}
\label{sec:con}

We have carried out a systematic search for possibilities to construct models 
in which we obtain preserved $\mathcal N=1$ SUSY, the SM gauge group, all the MSSM fields, three generation 
structures of quarks and leptons and full-rank Yukawa matrices. 

First, we have done that considering only the factorizable fluxes,
then we find that the Pati-Salam models with six pairs of Higgs multiplets can be 
constructed. 

An orbifold projection could eliminate exotic modes without 
changing three generation structures of quarks and leptons. It also 
can change the number of degenerate zero-modes if $Z_2$ act on the torus 
where flavor structures are generated, and we try to use that to generate 
three generation structures other than previous ones. With such an orbifold 
projection, (continuous) Wilson lines are not allowed and we have to break the gauge group 
using only magnetic fluxes. Thus, we have to consider five-block magnetic fluxes 
for which SUSY conditions become more severe and we find it impossible. 

We can consider nonfactorizable fluxes which cross over two tori. 
Considering nonfactorizable fluxes, more various model buildings 
become possible. With such fluxes we tried to construct the Pati-Salam models, 
and we find that there are 
infinite possibilities to give three generation models as a 
result of systematic research. We also 
find some particular values are allowed for the number of Higgs pairs 
if Higgs multiplets come from $\phi_3$, in contrast any numbers of 
Higgs pairs can be 
realized in the other case. Furthermore, the quite attractive fact is that 
we can construct one-pair Higgs models just like the MSSM. 
Finally, we tried to construct nonfactorizable flux models on 
$T^6/Z_2$ orbifolds. As a result we cannot find a flux configuration 
to give us the plausible model. 

In this paper we have found many possibilities to give the plausible models, 
which have the three generation quark and lepton multiplets and 
full-rank Yukawa matrices, 
other than the one studied in Ref.~\cite{Abe:2012fj}. Plausible flux 
configurations give only the Pati-Salam models, where we need introduce 
Wilson lines to break the gauge group. In the Pati-Salam models with only 
factorizable fluxes, 
Yukawa coupling possesses the flavor symmetry $\Delta(27)$\cite{Abe:2009vi}. 
It is interesting to study how such structures would be changed by nonfactorizable fluxes. 
Furthermore, introducing nonfactorizable fluxes has a close connection with 
the moduli stabilization. 
We will study these feature of nonfactorizable fluxes elsewhere.

The five-block flux models not needing Wilson lines cannot be realized in 
10D magnetized SYM theory. However, it would be straightforward to extend our 
study to SYM theories in a lower than ten-dimensional spacetime, or even to 
the mixture of SYM theories with a different dimensionality. For example, 
in type IIB orientifolds, our study will be adopted not only to the magnetized 
D9 branes, which are T-dual to intersecting D6 branes in the IIA 
side, but also to the D5-D9 \cite{DiVecchia:2011mf} and the D3-D7 brane configurations with magnetic fluxes in the extra dimensions. With such brane configurations, we might be able to 
build the five-block and more various models. 

As other generalizations, we can consider generalized torus in 
which off-diagonal elements of general complex structure $\Omega$ have 
nonzero values, or more complicated manifold, and that would give us 
a new flavor landscape. 

Finally, our models derived from supersymmetric Yang-Mills theory 
might be embedded in the matrix models 
\cite{Nishimura:2013moa}. 
In that case, it would give us a new motivation to study the magnetized 
supersymmetric Yang-Mills theory.

\subsection*{Acknowledgement}

H.A. was supported in part by the Grant-in-Aid for Scientific Research No.~25800158 from the Ministry of Education, Culture, Sports, Science 
and Technology (MEXT) in Japan. 
T.K. was supported in part by the Grant-in-Aid for Scientific Research 
No.~25400252 from the MEXT in Japan. 
H.O. was supported in part by the JSPS Grant-in-Aid for Scientific Research 
(S) No.~22224003, 
for young Scientists (B) No.~25800139 from the MEXT in Japan. 
K.S. was supported in part by a Grant-in-Aid for JSPS Fellows 
No.~25$\cdot$4968 and a Grant for Excellent Graduate
Schools from the MEXT in Japan. 
Y.T. was supported in part by a Grant for Excellent Graduate
Schools from the MEXT in Japan. 
Y.T. would like to thank T. Shoji and K. Oh for a correspondence.

\appendix

\section{Exotic modes with the pattern1 fluxes}
\label{sec:appa}

For the matter profile~(\ref{app.phis1}) caused by 
the type1 magnetic fluxes in Table.~\ref{tb:3models}, we see that 
$Z_2$-projection introduced in Sec.~\ref{sec:mag} which acts on the second and the third tori with 
the following projection operator, 
\begin{eqnarray}
P_{ab} &=& \left( 
\begin{array}{ccc}
-{\bm 1}_4 & 0 & 0 \\
0 & +{\bm 1}_2 & 0 \\
0 & 0 & +{\bm 1}_2 
\end{array} \right), 
\nonumber\label{app.opera}
\end{eqnarray}
removes most of the massless exotic modes $\Xi_{ab}^{(r)}$ 
and some of massless diagonal components $\Omega_{a}^{(r)}$. 
The matter contents on the orbifold $T^6/Z_2$ are found as 
\begin{eqnarray}
\phi_1^{{\cal I}_{ab}} &=& 
\left( 
\begin{array}{cc|c|cc}
\Omega_C^{(1)} & \Xi_{CC'}^{(1)} & 0 & 0 & 0 \\
\Xi_{C'C}^{(1)} & \Omega_{C'}^{(1)} & 0 & 0 & 0 \\ 
\hline 
0 & 0 & \Omega_L^{(1)} & H_u^K & H_d^K \\ 
\hline 
0 & 0 & 0 & \Omega_{R'}^{(1)} & \Xi_{R'R''}^{(1)} \\
0 & 0 & 0 & \Xi_{R''R'}^{(1)} & \Omega_{R''}^{(1)} 
\end{array}
\right), 
\nonumber \\
\phi_2^{{\cal I}_{ab}} &=& 
\left( 
\begin{array}{cc|c|cc}
0 & 0 & Q^I & 0 & 0 \\
0 & 0 & L^I & 0 & 0 \\
\hline 
0 & 0 & 0 & 0 & 0 \\
\hline 
0 & 0 & 0 & 0 & 0 \\
0 & 0 & 0 & 0 & 0 
\end{array}
\right), 
% \nonumber \\
\qquad 
\phi_3^{{\cal I}_{ab}} 
% &=& 
\ = \ 
\left( 
\begin{array}{cc|c|cc}
0 & 0 & 0 & 0 & 0 \\
0 & 0 & 0 & 0 & 0 \\
\hline 
0 & 0 & 0 & 0 & 0 \\
\hline 
U^J & N^J & 0 & 0 & 0 \\
D^J & E^J & 0 & 0 & 0 
\end{array}
\right), 
\nonumber
\end{eqnarray}
where $I,J=1,2,3$ and $K=1,\ldots,6$ 
label the generations as before. 
There still remain massless exotic modes $\Xi_{ab}^{(1)}$ 
for $a,b=C,C'$ and $a,b=R',R''$ with $a \ne b$ as well as 
open string moduli $\Omega_{a}^{(1)}$ for $a=C,C',L,R',R''$. 
That is one of the open problems in the $T^6/Z_2$ magnetized 
orbifold model. In Ref.~\cite{Abe:2012fj}, 
these exotic modes are assumed to become massive through 
some nonperturbative effects \cite{Ibanez:2006da} or higher-order corrections, 
so that they decouple from the low-energy physics.

\section{Yukawa matrices in a one-pair Higgs model}
\label{sec:appb}

We show an explicit form of Yukawa matrices in a one-pair Higgs model. 
%(General forms are studied in \cite{Antoniadis:2009bg,DeAngelis:2012jc})
Yukawa couplings of No.1 model in type1 are written by 
\begin{eqnarray*}
Y_{ijk} \propto \sum_{\vec{m}} \delta_{\vec{k}, \mathcal{M}_H^{-1} (\mathcal{M}_L \vec{i} + \mathcal{M}_R\vec{j}+ \mathcal{M}_L \vec{m})} \int d^2y \left[ e^{- \pi \vec{y}( \mathcal{M}_L \tilde{\Omega}_L + \mathcal{M}_R \tilde{\Omega}_R + \mathcal{M}_H \Omega) \cdot \vec{y}}
\cdot \vartheta \begin{bmatrix}\vec{\bold{K}}\\[5pt] 0 \end{bmatrix}(i\vec{\bold{Y}}|i\vec{\bold{Q}}) \right]
\end{eqnarray*}
where
\begin{gather*}
\tilde{\Omega}_{L}=\hat{\Omega}_{L} \cdot \Omega, \qquad \tilde{\Omega}_{R}=\hat{\Omega}_{R} \cdot \Omega\\
%\hat{\Omega}_{L}=\frac{1}{1+q_{ab}^2}\left(\begin{array}{cc}1-q_{ab}^2&-2q_{ab} \\ -2q_{ab}& -1+q_{ab}^2 \end{array}\right), \qquad \hat{\Omega}_{R}=\frac{1}{1+q_{ca}^2}\left(\begin{array}{cc}-1+q_{ca}^2&-2q_{ca} \\ -2q_{ca}& 1-q_{ca}^2 \end{array}\right)\\
\vec{\bold{K}} \equiv \left(\begin{array}{c} \vec{k}\\ (\vec{i}-\vec{j}+\vec{m}) \frac{\mathcal{M}_L(\mathcal{M}_L+\mathcal{M}_R)^{-1}\mathcal{M}_R}{\det\mathcal{M}_L \det\mathcal{M}_R} \end{array} \right)\\
\vec{\bold{Y}} \equiv \left(\begin{array}{c} (\mathcal{M}_L \tilde{\Omega}_{L} +\mathcal{M}_R \tilde{\Omega}_{R} +\mathcal{M}_H \Omega)\cdot \vec{y}\\ (\det\mathcal{M}_L \det \mathcal{M}_R) (\mathcal{M}_L \tilde{\Omega}_{L} (\mathcal{M}_L^{-1})^T - \mathcal{M}_R \tilde{\Omega}_{R} (\mathcal{M}_R^{-1})^T)\cdot \vec{y} ) \end{array} \right)\\
\vec{\bold{Q}} \equiv \left(\begin{array}{cc} \mathcal{M}_L \tilde{\Omega}_{L}+\mathcal{M}_R \tilde{\Omega}_{R} +\mathcal{M}_H\Omega&(\det\mathcal{M}_L\det\mathcal{M}_R) (\mathcal{M}_L \tilde{\Omega}_{L} \cdot(\mathcal{M}_L^{-1})^T - \mathcal{M}_R \tilde{\Omega}_{R}\cdot (\mathcal{M}_R^{-1})^T)\\
(\det\mathcal{M}_L\det\mathcal{M}_R)(\tilde{\Omega}_{L} - \tilde{\Omega}_{R})&(\det\mathcal{M}_L\det\mathcal{M}_R)^2 (\tilde{\Omega}_{L} \mathcal{M}_L^{-1} +\tilde{\Omega}_{R} \mathcal{M}_R^{-1})  \end{array} \right).
\end{gather*}
We define 
\begin{gather*}
\eta\left[(\vec{i}-\vec{j}+\vec{m}) \frac{\mathcal{M}_L(\mathcal{M}_L+\mathcal{M}_R)^{-1}\mathcal{M}_R}{\det\mathcal{M}_L \det\mathcal{M}_R}\right] \equiv
\vartheta \begin{bmatrix}\vec{\bold{K}}\\[5pt] 0 \end{bmatrix}(i\vec{\bold{Y}}|i\vec{\bold{Q}}),
\end{gather*}
and then $(1,1)$ element can be written using $\eta$;
\begin{gather*}
\lambda_{11}=\eta[(0,0)]+\eta[(0,\tfrac{1}{3})]+\eta[(0,\tfrac{2}{3})]+\eta[(\tfrac{1}{3},0)]+\eta[(\tfrac{1}{3},\tfrac{1}{3})]+\eta[(\tfrac{1}{3},\tfrac{2}{3})]+\eta[(\tfrac{2}{3},0)]+\eta[(\tfrac{2}{3},\tfrac{1}{3})]+\eta[(\tfrac{2}{3},\tfrac{2}{3})] .\notag
\end{gather*}
Similarly, other elements are written by 
\begin{align*}
\lambda_{12} &= \eta[(\tfrac{1}{9},\tfrac{1}{9})]+\eta[(\tfrac{1}{9},\tfrac{4}{9})]+\eta[(\tfrac{1}{9},\tfrac{7}{9})]+\eta[(\tfrac{4}{9},\tfrac{1}{9})]+\eta[(\tfrac{4}{9},\tfrac{4}{9})]+\eta[(\tfrac{4}{9},\tfrac{7}{9})]+\eta[(\tfrac{7}{9},\tfrac{1}{9})]+\eta[(\tfrac{7}{9},\tfrac{4}{9})]+\eta[(\tfrac{7}{9},\tfrac{7}{9})] \notag\\
\lambda_{13} &= \eta[(\tfrac{2}{9},\tfrac{2}{9})]+\eta[(\tfrac{2}{9},\tfrac{5}{9})]+\eta[(\tfrac{2}{9},\tfrac{8}{9})]+\eta[(\tfrac{5}{9},\tfrac{2}{9})]+\eta[(\tfrac{5}{9},\tfrac{5}{9})]+\eta[(\tfrac{5}{9},\tfrac{8}{9})]+\eta[(\tfrac{8}{9},\tfrac{2}{9})]+\eta[(\tfrac{8}{9},\tfrac{5}{9})]+\eta[(\tfrac{8}{9},\tfrac{8}{9})] \notag\\
\lambda_{21} &= \eta[(\tfrac{2}{9},\tfrac{1}{9})]+\eta[(\tfrac{2}{9},\tfrac{4}{9})]+\eta[(\tfrac{2}{9},\tfrac{7}{9})]+\eta[(\tfrac{5}{9},\tfrac{1}{9})]+\eta[(\tfrac{5}{9},\tfrac{4}{9})]+\eta[(\tfrac{5}{9},\tfrac{7}{9})]+\eta[(\tfrac{8}{9},\tfrac{1}{9})]+\eta[(\tfrac{8}{9},\tfrac{4}{9})]+\eta[(\tfrac{8}{9},\tfrac{7}{9})] \notag\\
\lambda_{22}&=\eta[(0,\tfrac{2}{9})]+\eta[(0,\tfrac{5}{9})]+\eta[(0,\tfrac{8}{9})]+\eta[(\tfrac{1}{3},\tfrac{2}{9})]+\eta[(\tfrac{1}{3},\tfrac{5}{9})]+\eta[(\tfrac{1}{3},\tfrac{8}{9})]+\eta[(\tfrac{2}{3},\tfrac{2}{9})]+\eta[(\tfrac{2}{3},\tfrac{5}{9})]+\eta[(\tfrac{2}{3},\tfrac{8}{9})] \notag\\
\lambda_{23}&=\eta[(\tfrac{1}{9},0)]+\eta[(\tfrac{1}{9},\tfrac{1}{3})]+\eta[(\tfrac{1}{9},\tfrac{2}{3})]+\eta[(\tfrac{4}{9},0)]+\eta[(\tfrac{4}{9},\tfrac{1}{3})]+\eta[(\tfrac{4}{9},\tfrac{2}{3})]+\eta[(\tfrac{7}{9},0)]+\eta[(\tfrac{7}{9},\tfrac{1}{3})]+\eta[(\tfrac{7}{9},\tfrac{2}{3})] \notag\\
\lambda_{31} &= \eta[(\tfrac{1}{9},\tfrac{2}{9})]+\eta[(\tfrac{1}{9},\tfrac{5}{9})]+\eta[(\tfrac{1}{9},\tfrac{8}{9})]+\eta[(\tfrac{4}{9},\tfrac{2}{9})]+\eta[(\tfrac{4}{9},\tfrac{5}{9})]+\eta[(\tfrac{4}{9},\tfrac{8}{9})]+\eta[(\tfrac{7}{9},\tfrac{2}{9})]+\eta[(\tfrac{7}{9},\tfrac{5}{9})]+\eta[(\tfrac{7}{9},\tfrac{8}{9})] \notag\\
\lambda_{32}&=\eta[(\tfrac{2}{9},0)]+\eta[(\tfrac{2}{9},\tfrac{1}{3})]+\eta[(\tfrac{2}{9},\tfrac{2}{3})]+\eta[(\tfrac{5}{9},0)]+\eta[(\tfrac{5}{9},\tfrac{1}{3})]+\eta[(\tfrac{5}{9},\tfrac{2}{3})]+\eta[(\tfrac{7}{9},0)]+\eta[(\tfrac{7}{9},\tfrac{1}{3})]+\eta[(\tfrac{7}{9},\tfrac{2}{3})] \notag\\
\lambda_{33}&=\eta[(0,\tfrac{1}{9})]+\eta[(0,\tfrac{4}{9})]+\eta[(0,\tfrac{7}{9})]+\eta[(\tfrac{1}{3},\tfrac{1}{9})]+\eta[(\tfrac{1}{3},\tfrac{4}{9})]+\eta[(\tfrac{1}{3},\tfrac{7}{9})]+\eta[(\tfrac{2}{3},\tfrac{1}{9})]+\eta[(\tfrac{2}{3},\tfrac{4}{9})]+\eta[(\tfrac{2}{3},\tfrac{7}{9})] ,\notag
\end{align*}
and total Yukawa matrices can be written as follows
\begin{gather*}
Y_{ijk} \propto \int d^2y \left[
e^{- \pi \vec{y} ( \mathcal{M}_L \tilde{\Omega}_{L} +\mathcal{M}_R \tilde{\Omega}_{R} + \mathcal{M}_H \Omega ) \vec{y}}
\left(\begin{array}{ccc} \lambda_{11}&\lambda_{12}&\lambda_{13}\\ \lambda_{21}&\lambda_{22}&\lambda_{23}\\  \lambda_{31}&\lambda_{32}&\lambda_{33} \end{array}\right)
\right].
\end{gather*}
The values of elements are different from each other, 
and we cannot find any non-Abelian flavor 
symmetry there.

\end{document}